\documentclass[structabstract]{aa}  
\usepackage{natbib}
\usepackage{array}
\usepackage{subfigure} 
\usepackage{amsmath}
\usepackage{graphicx}
\usepackage{txfonts}

\usepackage{aas_macros}
\allowdisplaybreaks[1]

\usepackage[normalem]{ulem}

\begin{document}

    \title {Simulating gamma-ray binaries with a relativistic extension of RAMSES}

    \author{A. Lamberts 
        \inst{1,2}
       \and  S. Fromang 
        \inst{3}
        \and  G. Dubus 
        \inst{1}
        \and R. Teyssier         
        \inst{3,4}
       }

     \institute{UJF-Grenoble 1 / CNRS-INSU, Institut de Plan\'{e}tologie et d\textquoteright Astrophysique de Grenoble (IPAG) UMR 5274, Grenoble, F-38041, France 
       \and
       Physics Department, University of Wisconsin-Milwaukee, Milwaukee WI 53211, USA
         \and
         Laboratoire AIM, CEA/DSM - CNRS - Universit\'e Paris 7, Irfu/Service d'Astrophysique, CEA-Saclay, 91191 Gif-sur-Yvette, France
         \and
         Institute for Theoretical Physics, University of Z\"urich, Winterthurestrasse 190, CH-8057 Z\"urich, Switzerland   }       
    \date{\today}

    \offprints{A. Lamberts: lambera@uwm.edu}


\titlerunning{Special relativity with RAMSES}
\authorrunning{A. Lamberts et al.}

\abstract
{$\gamma$-ray binaries are composed of a massive star and a rotation-powered pulsar with a highly relativistic wind. The collision between the winds from both objects creates a shock structure where particles are accelerated, which results in the observed high energy emission.}
 {We want to understand the impact of the relativistic nature of the pulsar wind on the structure and stability of the colliding wind region and highlight the differences with colliding winds from massive stars. We focus on how the structure evolves with increasing values of the Lorentz factor of the pulsar wind, keeping in mind that current simulations are unable to reach the expected values of pulsar wind Lorentz factors by orders of magnitude.}
{We use high resolution numerical simulations with a relativistic extension to the hydrodynamics code RAMSES we have developed. We use two-dimensional simulations, and focus on the region close to the binary, where orbital motion can be neglected. We use different values of the Lorentz factor of the pulsar wind, up to 16.}
{We determine analytic scaling relations between stellar wind collisions and $\gamma$-ray binaries. They provide the position of the contact discontinuity. The position of the shocks  strongly depends on the Lorentz factor. We find that the relativistic wind is more collimated than expected based on non-relativistic simulations. Beyond a certain distance, the shocked flow is accelerated to its initial velocity and follows adiabatic expansion. Finally, we provide guidance for extrapolation towards more realistic values of the Lorentz factor of the pulsar wind.}
{We extended the adaptive mesh refinement code RAMSES to relativistic hydrodynamics. This code is suited for the study of astrophysical objects such as pulsar wind nebulae, gamma-ray bursts or relativistic jets and will be part of the next public release of RAMSES.  Using this code we performed simulations of gamma-ray binaries up to $\Gamma_p=16$ and highlighted the limits and possibilities of current hydrodynamical models of gamma-ray binaries.  } 
\maketitle
\section{Introduction}

$\gamma$-ray binaries are composed of a massive star and a compact object.  They emit most of their power at GeV and TeV energies.  Up to now a  handful of such systems have been detected :  PSR B1259-63 (\citet{2005A&A...442....1A}), LS 5039 (\citet{2005Sci...309..746A}),  LSI +61\degr303  (\citet{2006Sci...312.1771A}) and more recently 1FGL J1018.6-5856 \citep{2012Sci...335..189F} and HESS J0632+057 \citep{2011ApJ...737L..11B}. 

 Pulsed radio emission  in PSR B1259-63 \citep{1992ApJ...387L..37J} indicates the compact object is a fast rotating pulsar, while the nature of the compact object in the other systems is less certain. In PSR B1259-63 the high-energy emission probably arises from the interaction between the relativistic wind from the pulsar and the wind from the companion star \citep{1994ApJ...433L..37T}. It displays extended radio emission showing a structure looking like a cometary tail whose orientation changes with orbital phases \citep{2011ApJ...732L..10M}.  It is interpreted as the evolution of a shock as the pulsar orbits around the massive star. The similarities in the variable high energy emission and the extended radio emission between PSR B1259-63 and the other detected $\gamma$-ray binaries suggest the wind collision scenario is at work in all the systems \citep{2006A&A...456..801D}.

$\gamma$-ray binaries share a common structure with colliding wind binaries composed of two massive stars.  Colliding stellar winds have been extensively studied in the past few years using numerical simulations \citep[see e.g.][]{2011ApJ...726..105P} that have highlighted the importance of various instabilities in the colliding wind region. The Kelvin-Helmholtz instability (KHI) can have a  strong impact, especially when the velocity difference between the winds is important. \citet{PaperII} (hereafter PaperII) show that it may destroy the expected large scale spiral structure. However, strong density gradients have a stabilizing effect. For strongly cooling winds, the shocked region is very narrow and the non-linear thin shell instability \citep{1994ApJ...428..186V} can strongly distort it  (\citep{PaperI}, hereafter PaperI).  For highly eccentric systems, the instability may grow only at periastron, where the higher density in the collision region leads to enhanced cooling \citep{2009MNRAS.396.1743P}. A very high resolution is required to model this instability and up to now large scale studies have been limited by their numerical cost. 

Both these instabilities create a turbulent colliding wind region, and may be responsible for the variability observed in stellar colliding wind binaries.   They can distort the positions of the shocks, thereby shifting the location of particle acceleration. Strong instabilities also lead to important mixing between the winds (PaperII). This may enhance thermalization of non-thermal particles and decrease their high energy emission.  All these effects may be at work in $\gamma$-ray binaries and affect their structure and emission properties \citep{2011A&A...535A..20B,2012A&A...544A..59B}.

The first numerical studies of the interaction between pulsar winds and their environment were performed in the hydrodynamical limit \citep{2003A&A...397..913V,2002A&A...387.1066B}. While providing good insights into the overall structure, a relativistic temperature or bulk motion of the plasma can have subtle effects. For instance, the normal and transverse velocities are coupled via the Lorentz factor even for a normal shock. The specific enthalpy of a relativistic plasma is also greater than for a cold plasma. A relativistic treatment is also necessary to estimate the impact of Doppler boosting on emission properties of the flow. Relativistic simulations have studied the morphology of the interaction between pulsar winds and  supernova remnants \citep{2003A&A...405..617B}, the interstellar medium \citep{2005A&A...434..189B} or the wind of a companion star \citep{2008MNRAS.387...63B}. Studies have  focused on the impact of magnetization and anisotropy in  the pulsar wind or orbital motion \citep{2007MNRAS.374..793V,2012MNRAS.419.3426B}.  They also  provide models for  non-thermal emission from pulsar wind nebulae \citep{2006A&A...453..621D,2008A&A...485..337V,2004MNRAS.349..779K}.

One of our goals is to determine the importance of relativistic effects both on the structure and stability of $\gamma$-ray binaries. We perform a set of 2D simulations of $\gamma$-ray binaries and we study the impact of the momentum flux ratio and the Lorentz factor of the pulsar wind on the colliding wind region.  We also determine the impact of the Kelvin-Helmholtz instability on the colliding wind region (\S\,\ref{sec:2D_gamma}).  We therefore extend the hydrodynamics code RAMSES to special relativistic hydrodynamics (\S\,\ref{sec:RAMSES_RHD}). We especially detail the implementation within the AMR scheme and present some tests (Appendix \S\,\ref{sec:num_details}) that validate this new code for relativistic hydrodynamics. We discuss how our results can be  extrapolated to more realistic values for the Lorentz factor of the pulsar wind (\S\,\ref{sec:discussion}) and conclude (\S\,\ref{sec:conclusion}).

\section{Relativistic hydrodynamics with RAMSES}\label{sec:RAMSES_RHD}
RAMSES \citep{Teyssier2002} is a numerical method for multidimensional astrophysical hydrodynamics and magnetohydrodynamics ~\citep{Fromang2006}. RAMSES uses an upwind second order Godunov method. The vector of conserved variables $\mathbf{U}$ is averaged over the volume of the cells and fluxes  $\mathbf{F}$ at cell interfaces are averaged in time. The updates in time are then given by

\begin{equation}
  \label{eq:godunov}
  \frac{\mathbf{U}^{n+1}_i-\mathbf{U}^{n}_i}{\Delta t}+\frac{\mathbf{F}^{n+1/2}_{i+1/2}-\mathbf{F}^{n+1/2}_{i-1/2}}{\Delta x}=0,
\end{equation}
where the subscript $i$ stands for the index of the cell and $n$ refers to the timestep.

RAMSES uses a Cartesian grid and allows the use of Adaptive Mesh Refinement (AMR), which enables to locally increase the resolution at a reasonable computational cost.

\subsection{Equations of relativistic hydrodynamics (RHD)}

Through all this paper the speed of light is $c \equiv 1$. In the frame of the laboratory the 3D-RHD equations for an ideal fluid can be written as a system of conservation equations \citep{Landau}.
\begin{eqnarray}\label{eq:RHD}
\frac{\partial{D}}{\partial{t}}+\frac{\partial{(Dv_k)}}{\partial{x_k}} &=&0\\ \label{eq:rel1}
\frac{\partial{M_j}}{\partial{t}}+\frac{\partial{(M_jv_k+P\delta^{j,k})}}{\partial{x_k}} &=&0\\ \label{eq:rel2}
\frac{\partial{E}}{\partial{t}}+\frac{\partial{(E+P)v_k}}{\partial{x_k}} &=&0 \label{eq:rel3},
\end{eqnarray}

These equations can be expressed in compact form
\begin{equation}\label{eq:cons}
\frac{\partial{\mathbf{U}}}{\partial{t}}+ \sum_k^N \frac{\partial{\mathbf{F_k}}}{\partial{x_k}}=0,
\end{equation}
where the vector of conservative variables is given by 
\begin{equation}\label{eq:cons_prim}
 \mathbf{U}=
 \begin{pmatrix} 
D \\ 
M_j\\
E 
\end{pmatrix}
=
\begin{pmatrix}
\Gamma \rho \\ 
\Gamma^2 \rho h v_j\\ 
\Gamma^2\rho h -P
\end{pmatrix}
.
\end{equation}
The fluxes $\mathbf{F_k}$ along each of the $N$ directions are given by
\begin{equation}
  \label{eq:flux_rhd}
\mathbf{F_k}= 
 \begin{pmatrix}
   \rho \Gamma v_k\\
   \rho h \Gamma^2v_jv_k+P\delta^{jk}\\
   \rho h\Gamma^2v_k
   \end{pmatrix}
.
\end{equation}
D is the density, $\mathbf{M}$ the momentum density and E the energy density in the frame of the laboratory.  The subscripts $j,k$ stand for the dimensions, $\delta^{jk}$  is the Kronecker symbol. $h$ is the specific enthalpy given by
\begin{equation}\label{eq:h}
h=h(P,\rho)=1+\frac{\gamma}{\gamma-1}\frac{P}{\rho}=\frac{e+P}{\rho},
\end{equation}
and $\rho$ is the proper mass density, $v_j$ is the fluid three-velocity, $P$ is the gas pressure and $e$ is the sum of the internal energy and rest mass energy of the fluid. The Lorentz factor is given by
\begin{equation}\label{eq:Lorentz}
\Gamma=\frac{1}{\sqrt{1-v^2}}.
\end{equation}
A passive scalar $s$ can be included in the simulations  using $S=s\rho \Gamma$ as the conserved variable and $F=\rho s v \Gamma$ to compute its flux.

An equation of state closes the system of equations \ref{eq:RHD}-\ref{eq:rel3}. The most commonly used is the so-called classical equation of state, where the rest mass energy is removed from the total internal energy $e$:
\begin{equation}\label{eq:EOS_ID}
P=(\gamma -1)(e-\rho),
\end{equation}
where $\gamma$ is the ratio of specific heats (or adiabatic index) which is constant and should not be confused with the Lorentz factor $\Gamma$.  In the non-relativistic limit $\gamma=5/3$, in the ultrarelativistic limit $\gamma=4/3$.  The sound speed is given by 
\begin{equation}\label{eq:cs_rel}
c_s^2\equiv \left(\frac{\partial P}{\partial e}\right)_s =\gamma\frac{P}{h\rho}.
\end{equation}
This leads to $c_s<1/3$ in the ultrarelativistic limit and $c_s<2/3$ in the non-relativistic limit. The kinetic theory of relativistic gases \citep{Synge} shows that the ratio of specific heats cannot be kept constant when passing from non-relativistic to highly relativistic temperatures and provides an equation of state which is valid for all temperatures.  Due to the complexity of the expressions relating different thermodynamical quantities, approximations have been developed for the  implementation in numerical schemes \citep{1996MNRAS.278..586F,2005ApJS..160..199M,2006ApJS..166..410R,2007MNRAS.378.1118M}. This has not been implemented yet in the relativistic extension of RAMSES that we present.

The RHD equations have a similar structure to the Euler equations and can be solved following the same  numerical method, making only localized changes.  Still, the strong coupling of the equations through the presence of the Lorentz factor and the enthalpy makes the resolution of the equations more complex.  Moreover, in relativistic dynamics,  velocities are bounded by the speed of light, resulting in an important numerical constraint.   These changes  particularly affect the determination of the primitive variables, the development of a second order scheme (\S,\ref{sec:second_order}), and the AMR framework (\S\,\ref{sec:AMR}).

\subsection {Second order numerical scheme}\label{sec:second_order}
Several steps in the algorithm require the use of the primitive variables 
\begin{equation}
  \label{eq:prim}
\mathbf{q}= 
 \begin{pmatrix}
   \rho \\
   v_j\\
   P
   \end{pmatrix}.
\end{equation}
In classical hydrodynamics, analytic relations allow a straightforward conversion from conservative to primitive variables. In RHD, there are no such relations, and several methods have been developed to recover the primitive variables  \citep[see e.g.][]{Aloy1999,2006ApJ...641..626N,2006ApJS..166..410R}. We use the method described in \citet{2007MNRAS.378.1118M}, that uses a Newton-Raphson algorithm to solve an equation on $W'=W-D=\rho h\Gamma^2-\rho\Gamma$. The details of the computation are given in Appendix\,\ref{sec:app_ctoprim}.

The determination of the fluxes $\mathbf{F}$ cells in Eq.\ref{eq:godunov} involves solving the Riemann problem at the interface between cells. Different solvers have been developed and some have been extended to RHD. We have implemented the relativistic HLL and HLLC solver in RAMSES. For the HLL Riemann solver, the expression of the Godunov flux is identical to the non relativistic expression \citep{Schneider1993}. However,  the computation of the maximal  wavespeeds propagating towards the left and right is different. In Newtonian hydrodynamics the wavespeed is the  sum of the sound speed and the advection speed of the flow. The relativistic composition of velocities couples the velocity of the flow parallel  and perpendicular  to the direction of spatial derivation and all components of the velocity need to be taken into \citep{2002A&A...390.1177D}. The development of a HLLC solver for RHD was done by \citet{2005MNRAS.364..126M}, whose method we closely follow.

To reach a high enough accuracy in scientific applications, second order schemes are however necessary. In such cases, fluxes are determined half a timestep ahead of the current timestep. We have implemented  the Monotonic Upstream Scheme for Conservation Laws (MUSCL)  and Piecewise Linear (PLM) methods.  In both schemes, the update is performed on the primitive variables, which are determined by a Taylor expansion 

\begin{equation}
  \label{eq:taylor}
  \mathbf{q}^{n+1/2}_{i\pm1/2, L,R}=\mathbf{q}^n_i\pm \Delta\mathbf{q}^n_i +\frac{d\mathbf{q}_i^n}{dt}\frac{\Delta t}{2},
\end{equation}
The subscript $L$ and $R$ respectively stand for the left and right side of a cell boundary,  $\Delta \mathbf{q}_i^n$ are the slopes of the variables in the cell, computed using a slope limiter, which is not affected by special relativity. $d\mathbf{q}_i^n$ is the temporal variation, and is given by 
\begin{equation}
  \label{eq:evol_prim}
\frac{\partial{\mathbf{q}}}{\partial{t}}+ \sum_k^N \mathbf{A}\frac{\partial{\mathbf{q}}}{\partial{x}} =0.
\end{equation}
where $\mathbf{A}=\partial\mathbf{F(q)}/\partial\mathbf{U(q)}$ is the Jacobian matrix of the system.  In classical hydrodynamics, flows in a given direction are not affected by motions perpendicular to that direction. In RHD all spatial directions are coupled through the Lorentz transformation: $\mathbf{A}$ is  $5\times5$ matrix even for 1D flows \citep{Pons_2000}. The value of $\mathbf{A}$ and its equivalents along the $y$ and $z$ direction are given in Appendix\,\ref{sec:app_jac}.

In a MUSCL-Hancock scheme \citep{vanLeer1979101}, one directly uses
\begin{equation}
  \label{eq:evolution}
  d \mathbf{q}=-\mathbf{A}\frac{\partial \mathbf{q}}{\partial x} dt,
\end{equation}
 to reconstruct the variables. In RAMSES, we have found that this reconstruction gives satisfactory results with the \textit{minmod} \citep{1986AnRFM..18..337R} slope limiter but fails with the \textit{moncen} limiter \citep{1977JCoPh..23..263V}. Therefore we have implemented  the PLM method, which works with both limiters.

In the  Piecewise Linear Method \citep{Collela_1990}, one rewrites Eq.\,\ref{eq:evol_prim} 
\begin{equation}
  \label{eq:PLM}
\frac{\partial{\mathbf{q}}}{\partial{t}}+ \sum_k^N \mathbf{L}^{\alpha}\lambda^{\alpha}\mathbf{R}^{\alpha}\frac{\partial{\mathbf{q}}}{\partial{x}} =0
\end{equation}
with  $\lambda^{\alpha}$ the eigenvalues  and  $(\mathbf{L}^{\alpha},\mathbf{R}^{\alpha})$ the eigenvectors of the Jacobian matrix.  In this case, the slopes are  projected on the characteristic variables, using the left eigenvectors. The waves that cannot reach the interface between cells in the given timestep are filtered out and do not contribute to the interface states. One then recovers the primitive variables, multiplying by the right eigenvectors. The final interface states are then computed by adding  the contributions of all the selected waves.

\begin{eqnarray}
  \label{eq:plmde}
  \mathbf{q}^{n+1/2}_{i,R}&=&\mathbf{q}^n_i + \sum_{\lambda_i^{\alpha}<0} \left(\frac{1}{2}\left(-1-\lambda_i^{\alpha}\frac{dt}{dx}\right)\mathbf{L_i}^{\alpha} \partial\mathbf{q}_i\right) \cdot \mathbf R_i^{\alpha}\\
 \mathbf{q}^{n+1/2}_{i,L}&=&\mathbf{q}^n_i + \sum_{\lambda_i^{\alpha}>0} \left(\frac{1}{2}\left(1-\lambda_i^{\alpha}\frac{dt}{dx}\right)\mathbf{L_i}^{\alpha} \partial\mathbf{q}_i\right) \cdot \mathbf R_i^{\alpha}.
\end{eqnarray}
The full set of eigenvalues and eigenvectors is given in Appendix \ref{sec:app_jac}.

As all directions are reconstructed separately, although each component is subluminal, nothing guarantees that the norm of the total velocity remains subluminal. When a superluminal velocity is obtained, one possibility is to reconstruct $\Gamma\mathbf{v}$ and renormalise the velocity when necessary \citep{Aloy1999} or to reconstruct the Lorentz factor independently \citep{2008ApJS..176..467W}. When superluminous velocities are obtained, we choose to switch to first order reconstruction ($\mathbf{q}_{i\pm1/2,L,R}=\mathbf{q}_i)$ in all directions, for all variables. No switch occurred in our test simulation of a relativistic jet (see \S\,\ref{sec:jet}). It typically occurs once every  $10^7$ updates in our gamma-ray binary simulations with a Lorentz factor of 7, probably due to the much larger region covered by the high velocity flow and its very low pressure (see \S \ref{sec:geometry}).

\begin{figure}[h]
  \centering
  \includegraphics[width = .4\textwidth ]{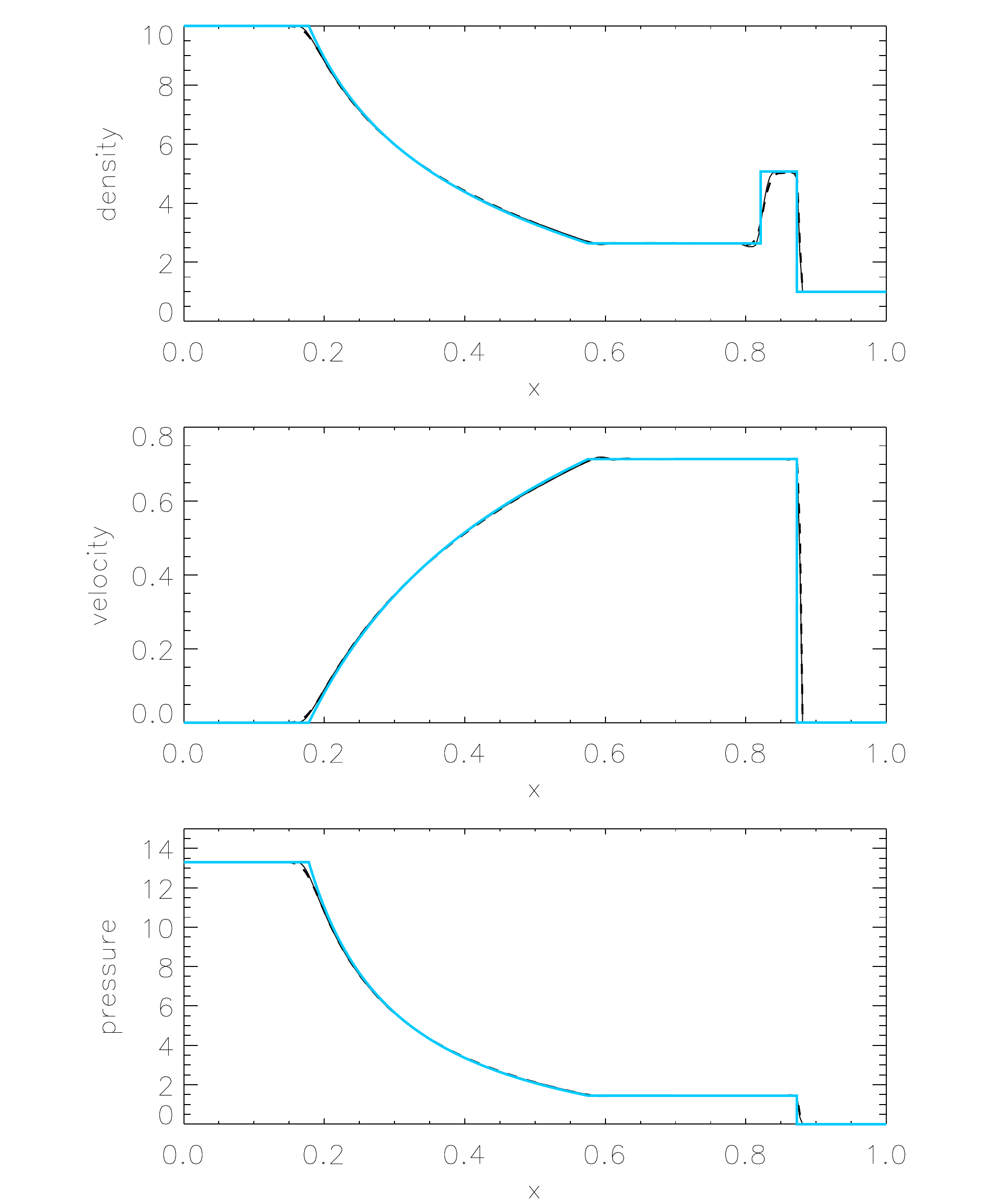}
  \caption{Density, velocity and pressure in the laboratory frame at $t=0.45$. The thin solid line shows the PLM method, the dashed line the MUSCL metod. Both methods give very similar results. The analytic solution \citep{2006JFM...562..223G} is given in blue.  The resolution is $n_ x=256$. }
  \label{fig:muscl_plmde}
\end{figure}

\begin{figure}[h]
  \centering
  \includegraphics[width = .4\textwidth ]{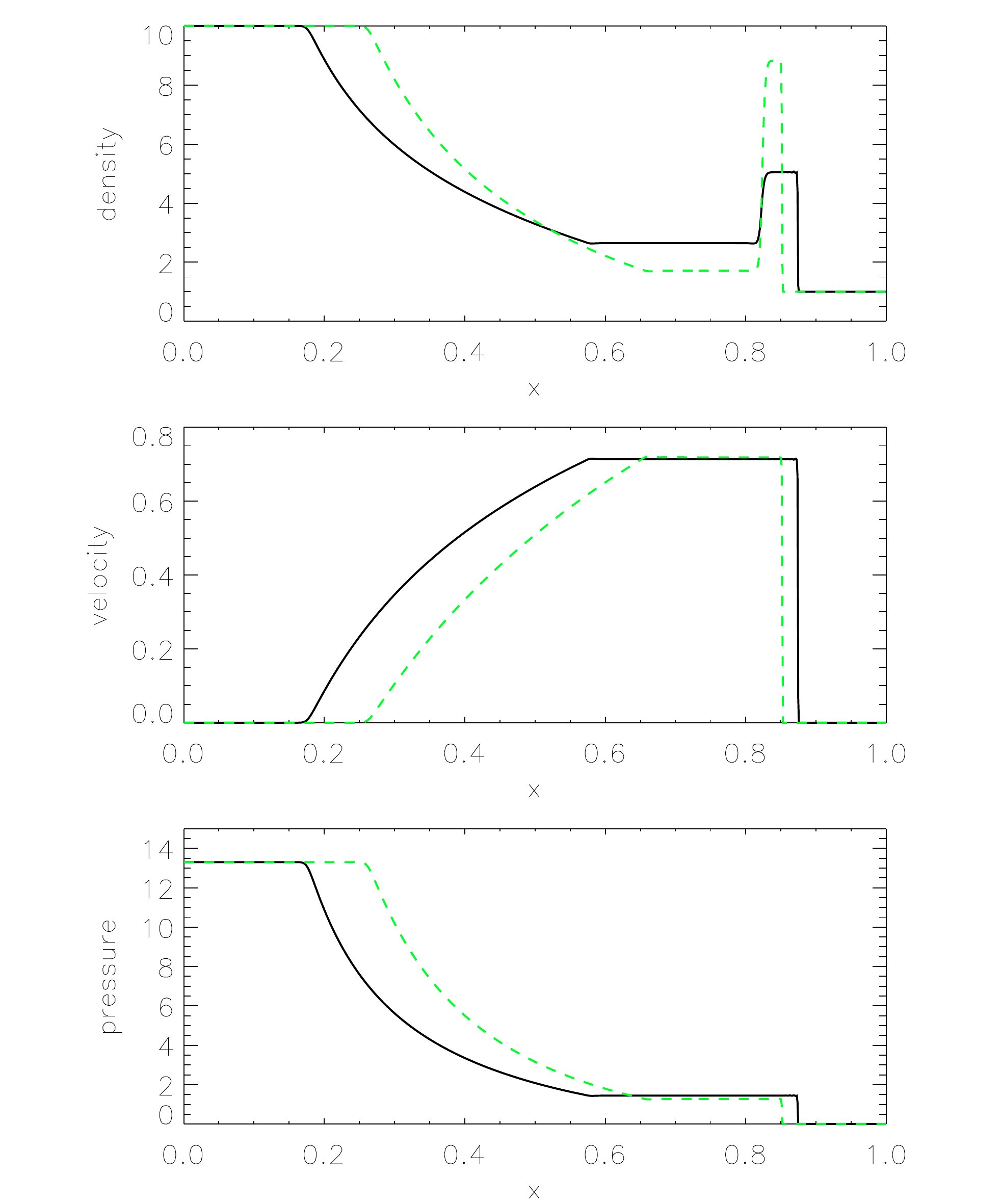}
  \caption{ Density, velocity and pressure in the laboratory frame at $t=0.45$ for tests with different adiabatic indices. The black solid line shows $\gamma=5/3$, the dahsed green line shows $\gamma=4/3$. The resolution is $n_ x=256$. }
  \label{fig:ad_index}
\end{figure}

Fig.\,\ref{fig:muscl_plmde} shows the result of a shock tube test for the two different methods. Both methods give comparable results and show the code works well on a uniform grid. In this  test \citep{2003LRR.....6....7M} $\rho_L=10$, $\rho_{R}=1$, $P_L=13.3$ and $P_{R}=10^{-6}$ and there is no initial velocity.  The flow is only mildly relativistic in the dynamical sense but $e\gg \rho$ in the right state, resulting in a thermodynamically highly relativistic flow. Fig.\ref{fig:ad_index} illustrates the differences between a non-relativistic adiabatic index ($\gamma=5/3$) and the ultrarelativistic adiabatic index ($\gamma=4/3)$.  The compression ratio in the shocked region is higher than for Newtonian hydrodynamics, especially when the gas is thermodynamically ultrarelativistic. The results compare very well with Fig. 2 in \citet{2006ApJS..166..410R}. 

\subsection{Implementation within the AMR structure}\label{sec:AMR}

Mesh refinement is an important tool  when modeling a wide range of spatial structures. Several methods exist and have been successfully associated with RHD schemes : static mesh refinement \citep{2011ApJS..193....6B}, block-based Adaptive Mesh Refinement \citep{2002ApJ...572..713H,2006ApJS..164..255Z,2012ApJS..198....7M} or tree-based AMR such as in \citet{Keppens2012718}.  Moving meshes offer another possibility to enable locally increased resolution \citep{2011ApJS..197...15D}.

In RAMSES, cells are related in a recursive tree-structure \citep{1997ApJS..111...73K} and gathered together in octs of $2^N$ cells ($N$ is the number of dimensions), which share the same parent cell.  When creating new refined cells or computing the fluxes at the interface between two levels, the conserved variables need to be interpolated from level $l$ to level $l+1$. The interpolation is done at second order, using a \textit{minmod} limiter for the linear reconstruction. Conversely, as hydrodynamical updates are only performed at the highest level of refinement, variables need to be determined at lower levels by computing the average over the whole oct. 

In RHD, the restriction step may lead to failures, when the energy is strongly dominated by the kinetic energy.  In RHD, positive pressure and subluminal velocity is guaranteed when $E^2>M^2+D^2$ (\citealt{2005MNRAS.364..126M}, see Appendix \ref{sec:num_details} for the demonstration). Although cells at a given level satisfy this condition, nothing guarantees that $E_{oct}^2>M_{oct}^2+D_{oct}^2$ where the subscript $oct$ means variables are summed over an oct.  The resulting state can be non-physical.    We found that this problem can be bypassed by performing the reconstruction on the specific internal energy $\epsilon$ ({\em i.e.} the temperature) rather than on the total energy using
\begin{equation}
  \label{eq:rhd_amr}
  \epsilon =\frac{P}{\rho}\frac{1}{\gamma-1},
\end{equation}
where $P$ and $\rho$ are computed with the Newton-Raphson scheme used to computed the primitive variables. After the restriction one recovers the total energy using
\begin{equation}
  E=\Gamma^2\rho h-P=Dh\Gamma-(\gamma-1)\frac{D}{\Gamma}\epsilon,
\end{equation}
where
\begin{equation}
  h=1+\gamma \epsilon \qquad \textrm{and} \qquad \Gamma=\sqrt{\frac{M^2}{D^2h^2}+1}.
\end{equation}

This method makes the numerical scheme non-conservative but guarantees the pressure is  positive and the speed is subluminal. 

\begin{figure}[h]
  \centering
  \includegraphics[width = .4\textwidth ]{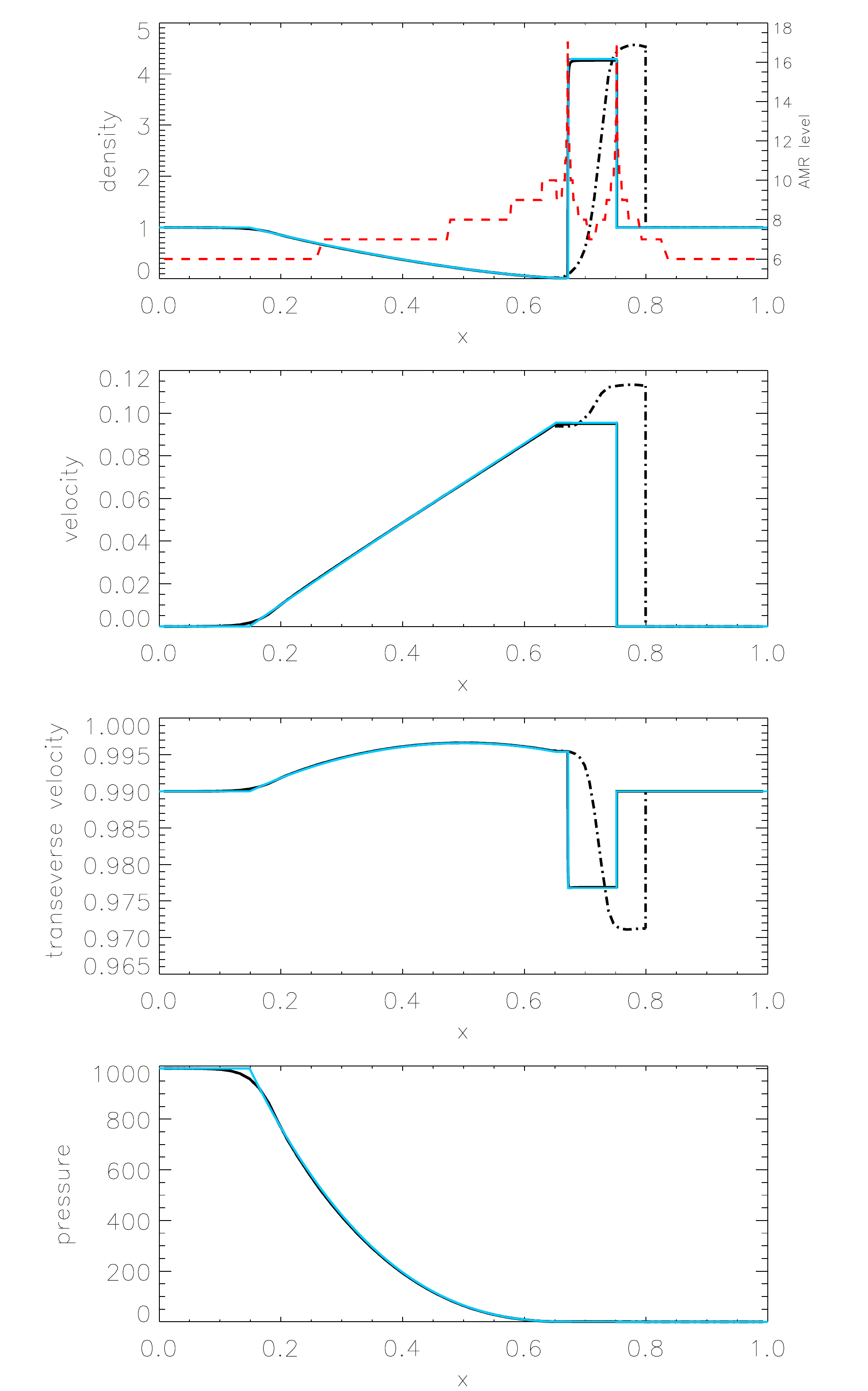}
  \caption{Density, parallel velocity, transverse velocity, and pressure in the laboratory frame for two different simulations. In both cases, the coarse grid is set by $n_x=64$ with 11 levels of refinement based on pressure gradients. The solid line shows a simulation where additional refinements occur based on Lorentz factor gradients. The result is superposed to  the analytic solution \citep{2006JFM...562..223G}  given in blue. The red dashed line in the upper panel shows the refinement levels, based on the Lorentz factor gradients.}
  \label{fig:AMR_Lor}
\end{figure}

Refinement is often based on gradients in the flow variables, either the density, pressure or velocity. In highly relativistic flows, the velocity does not vary significantly and the variations in the Lorentz factor may not be captured properly, strongly altering the dynamics of highly relativistic flows. To avoid this, we have implemented refinement based on the gradients of the Lorentz factor. Fig.\,\ref{fig:AMR_Lor} shows two simulations of a shock tube test.  We reproduce the test by  \citet{Pons_2000}, with $\rho_L=\rho_{R}=1$, $P_L=10^3$ and $P_{R}=10^{-2}$, $v_{x,L}=v_{x,R}=0$, $v_{y,L}=v_{y,R}=0.99$ initially.  This is a very stringent test, with a maximal Lorentz factor of 120. In the first case (dashed line), refinements are only based on pressure gradients and do not properly capture the contact discontinuity and positions of the shock. In the other simulation (solid line), refinement is also based on the Lorentz factor. In this case, the simulation is in very good agreement with the analytic result (blue line). Similar accuracy can also be found by refining according to both the pressure and density gradients. However, it increases the computational cost by  $\simeq 15\%$ in this test.  This tests validates the implementation within the AMR framework, which shows to be well adapted for shocks in highly relativistic flows.   We estimated the level of non-conservation by measuring the variation of the total density, momentum and energy (in the laboratory frame) at the end of the test. We found relative errors of $3\times 10^{-9},\,9\times 10^{-5}$ and $-6\times 10^{-8}$, respectively. 

This method is currently implemented in RAMSES and succesfully passes all the commonly used numerical tests. A few of them are detailed in Appendix\,\ref{sec:tests}.

\section{2D simulations of $\gamma$-ray binaries}\label{sec:2D_gamma}

Pulsar winds have a Lorentz factor of about $10^4-10^6$ \citep[see e.g.][for a review]{2009ASSL..357..421K}, which is far beyond the reach of current computer power and numerical schemes. Most present multidimensional simulations model Lorentz factors of at most a few 10. Modeling higher Lorentz factors, especially  in low density and/or low pressure environments require highly sophisticated numerical methods,  particularly in 2D and 3D, as spatial directions are strongly coupled. Up to now, there has been no extensive study on the impact of relativistic effects regarding the interaction of a pulsar wind with its surroundings. 

The following simulations are demanding, not only because they need a high resolution, but also because of the different timescales involved. The longest dynamical time is set by the speed of the stellar wind while the timestep in the simulation is limited by the speed of the pulsar wind. As $v_p/v_*\simeq 100$, simulating a $\gamma$-ray binaries takes roughly hundred times longer than simulating colliding stellar winds.  The simulations in this section typically took between 10 000 and 15 000 CPU hours to complete.

\subsection{Collision with supersonic relativistic winds}

We compare the interaction between a pulsar wind and a stellar wind to the collision between two stellar winds detailed in PaperI. Similarly to the non-relativistic case, one can determine the position of the contact discontinuity  between the winds. The standoff point, at the intersection between the line-of-centres and the contact discontinuity is given by the equation of momentum fluxes
\begin{equation}
  \label{eq:mom_eq_rel}
  \rho_p h_p \Gamma_p^2 v_p^2+P_p=\rho_s v_s^2+P_s,
\end{equation}
where the subscript $s$ represents the variables in the stellar wind, the variables with subscript $p$ refer to the pulsar. We neglect thermal pressure in the winds, which gives
\begin{equation}
  \rho_p \Gamma_p^2 v_p^2=\rho_s v_s^2.
\end{equation}
The mass loss rate $\dot{M}\equiv 4\pi r^2 \Gamma \rho v$ is constant so the location of standoff point, where Eq.~\ref{eq:mom_eq_rel} is realized, is set by the dimensionless ratio
\begin{equation}
  \label{eq:eta_rel}
 \eta_{\rm rel}=\frac{\dot{M}_p\Gamma_p v_p}{\dot{M_s}v_s}=\eta_{\rm cl} \Gamma
\end{equation}
where $\eta_{\rm cl}$  \citep{Stevens:1992on} is the usual definition of the momentum flux ratio of the winds in the non-relativistic limit. These equations suggest that we can expect a similar structure than in  colliding stellar winds, provided we define the momentum flux ratio of the winds using Eq.~\ref{eq:eta_rel}.

\subsection{Numerical setup\label{num}}
We perform 2D numerical simulations of the interaction in a region extending up to four times the binary separation $a$. Our 2D setup is cylindrical, as described in paper I. We neglected orbital motion in our simulations to enable comparisons with the analytic estimates and clarify the differences with our previous non-relativistic results (paper I). Orbital motion can be neglected without affecting the dynamics on scales much smaller than the spiral stepsize $S\ga v_s P_{\rm orb}$ (paper II). For LS 5039, the $\gamma$-ray binary with the shortest known period, the step of the spiral is at least $S\simeq 4$ AU, which about 20 times the binary separation \citep{2012A&A...548A.103M}.

 As we want to study the impact of relativistic effects, we do not set the pulsar wind velocity to $\simeq c$ but keep it as a free parameter. We set $\eta_{\rm rel}$ and $v_p$ and derive the pulsar's mass loss rate
\begin{equation}
  \label{eq:pulsar_mass_loss}
  \dot{M_p}=\eta_{\rm rel}\frac{\dot{M}_sv_s}{v_p\Gamma_p}.
\end{equation}

As for the classical case, the winds are initialised in 'masks' following the method described in \citet{Lemaster:2007sl}. Density is determined by mass conservation in the winds, the velocity is set to the terminal velocity. As in stellar wind simulations, we set the Mach number $\mathcal{M}=v/c_s=30$ at the distance $r=a$. We introduce a passive scalar, which is set to 1 in the stellar wind and set to 0 in the pulsar wind.

To ease comparisons between the different simulations, we set the adiabatic index $\gamma$ to 5/3 and keep the parameters of the stellar wind identical in all simulations. We set $\dot{M}_s=10^{-7}$ M$_{\odot}$ yr$^{-1}$ and $v_s=3000\,$km\,s$^{-1}$, corresponding to typical values for winds from early type stars \citep{2008A&ARv..16..209P}.  At this stage, we use the HLL Riemann solver to numerically quench the development of the Kelvin-Helmholtz instability at the contact discontinuity between the winds. We use the MUSCL scheme, combined with the \textit{minmod} slope limiter. In most simulations, the binary is at the center of a region of size $l_{\rm box}=8$\,$a$, the coarse grid is set by $n_x=64$, the refinement is based on density and Lorentz factor gradients.

\begin{figure}[!h]
  \centering
  \includegraphics[width = .4\textwidth ]{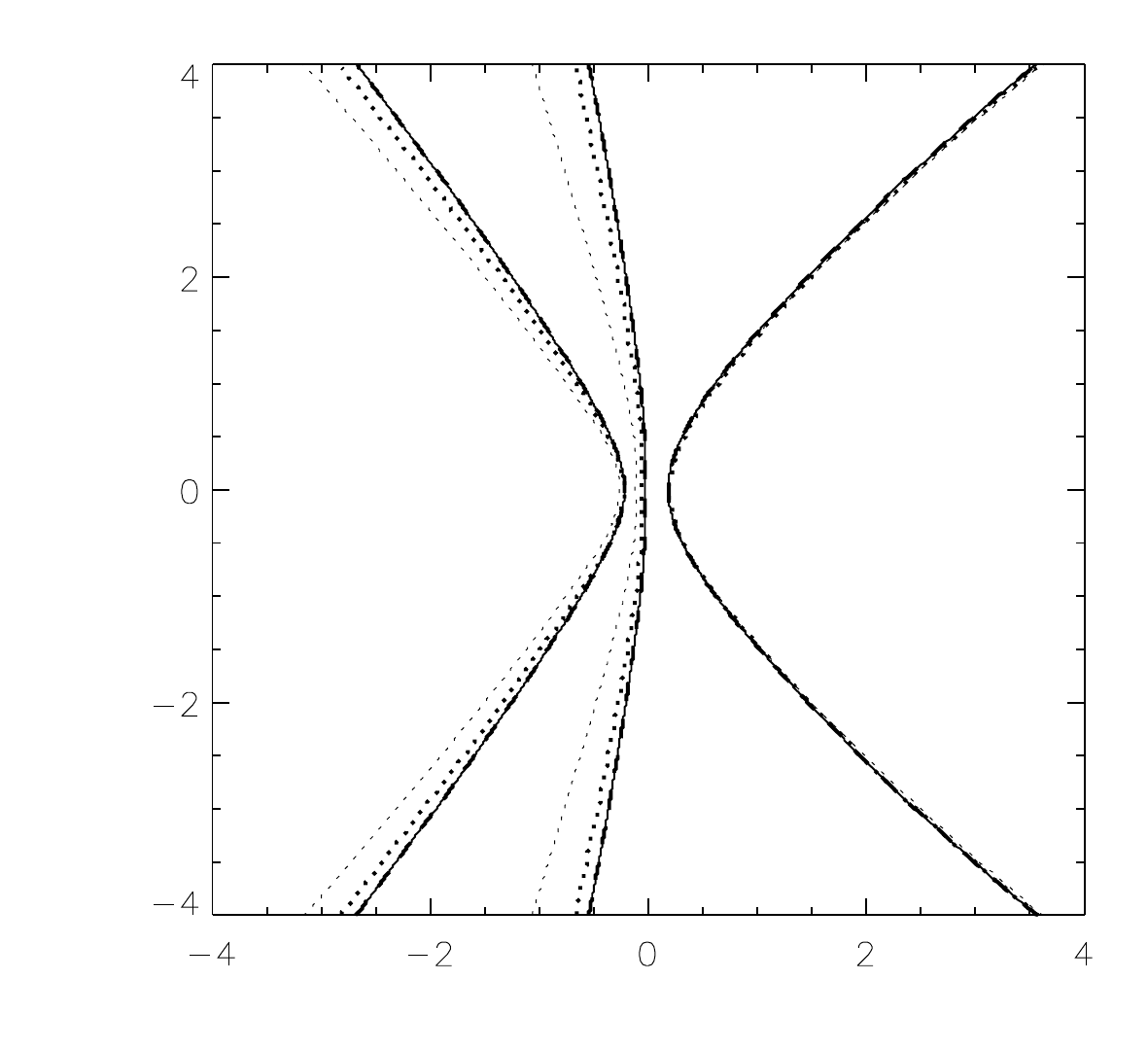}  
   \caption{Position of both shocks and the contact discontinuity for $\eta_{\rm rel}=1$, $\beta_p=0.99$ for increasing resolution $l_{\rm max}=$ 10 (thin dotted line), 11 (thick dotted line), 12 (dashed line), 13 (solid line). The last two resolutions give the same result.}\label{fig:relat_convergence}
\end{figure}

We  have performed test simulations for $\eta_{\rm rel}=1$  at different resolutions to determine the level of refinement required to reach numerical accuracy.  The Sod test problem (\S\ref{sec:1D_sod}) suggest that the higher the Lorentz factor of the flow, the higher the resolution needed to obtain a satisfactory simulation.   We consider a simulation provides satisfactory results when the positions of the discontinuities do not vary with increasing resolution. The position of the contact discontinuity is set by finding where the passive scalar equals 0.5.  The positions of the two shocks are given by the minimal and maximal value of the derivative of the pressure.  Fig.\,\ref{fig:relat_convergence} shows the positions of the discontinuities for different levels of resolution, for $\beta_p=v_p/c=0.99$. This corresponds to $\Gamma_p=7.08$. For $l_{\rm max}\geq 12$, the positions do not vary anymore.  In comparison, $l_{\rm max}=9$ is sufficient for the non-relativistic wind in Fig.\,\ref{fig:relat_convergence}. A much higher resolution is required to properly model the interaction with a relativistic flow.

\subsection{Geometry of the colliding wind region}\label{sec:geometry}

\begin{figure*}
  \centering
  \includegraphics[width = .3\textwidth ]{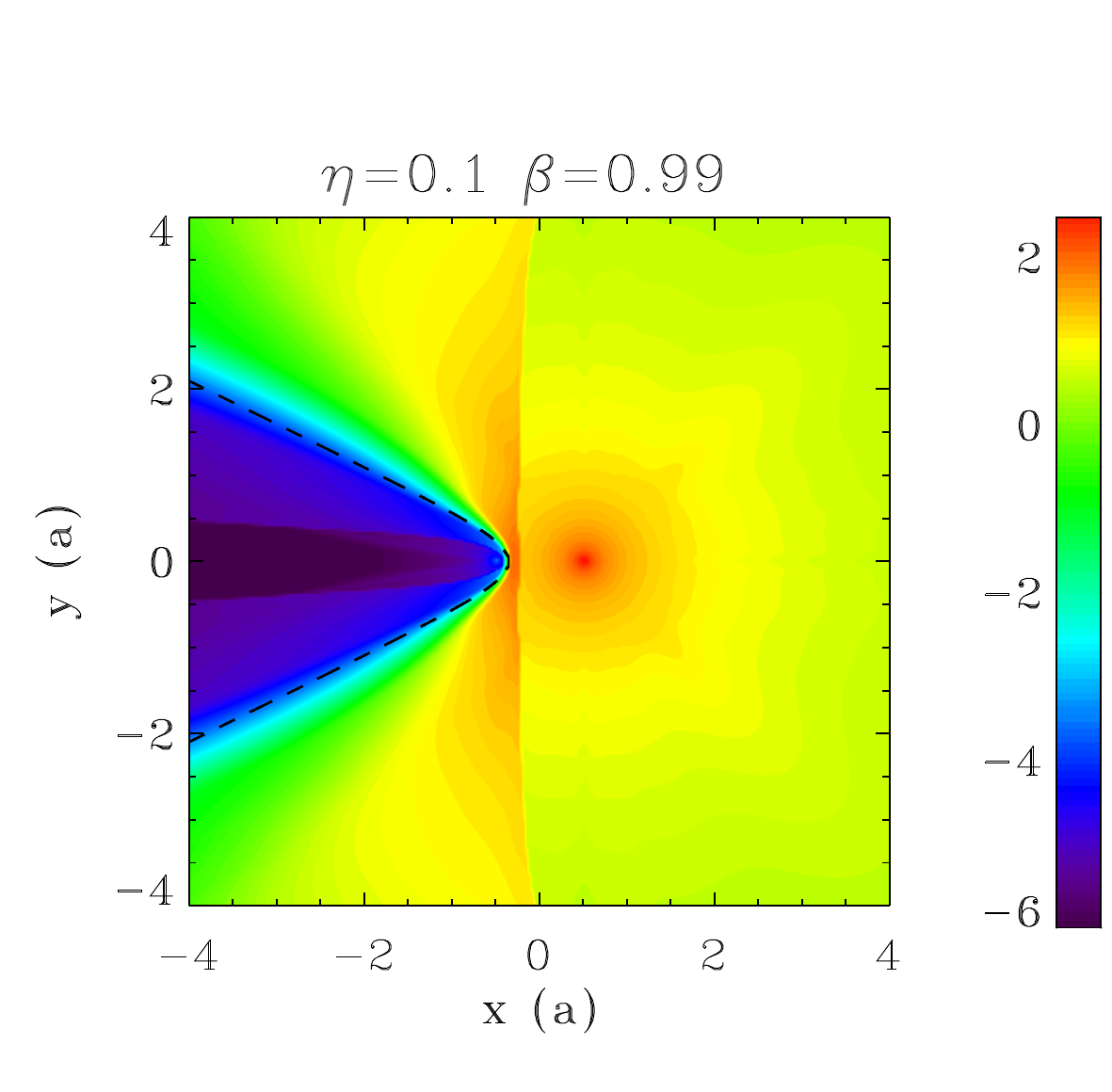}
  \includegraphics[width = .3\textwidth ]{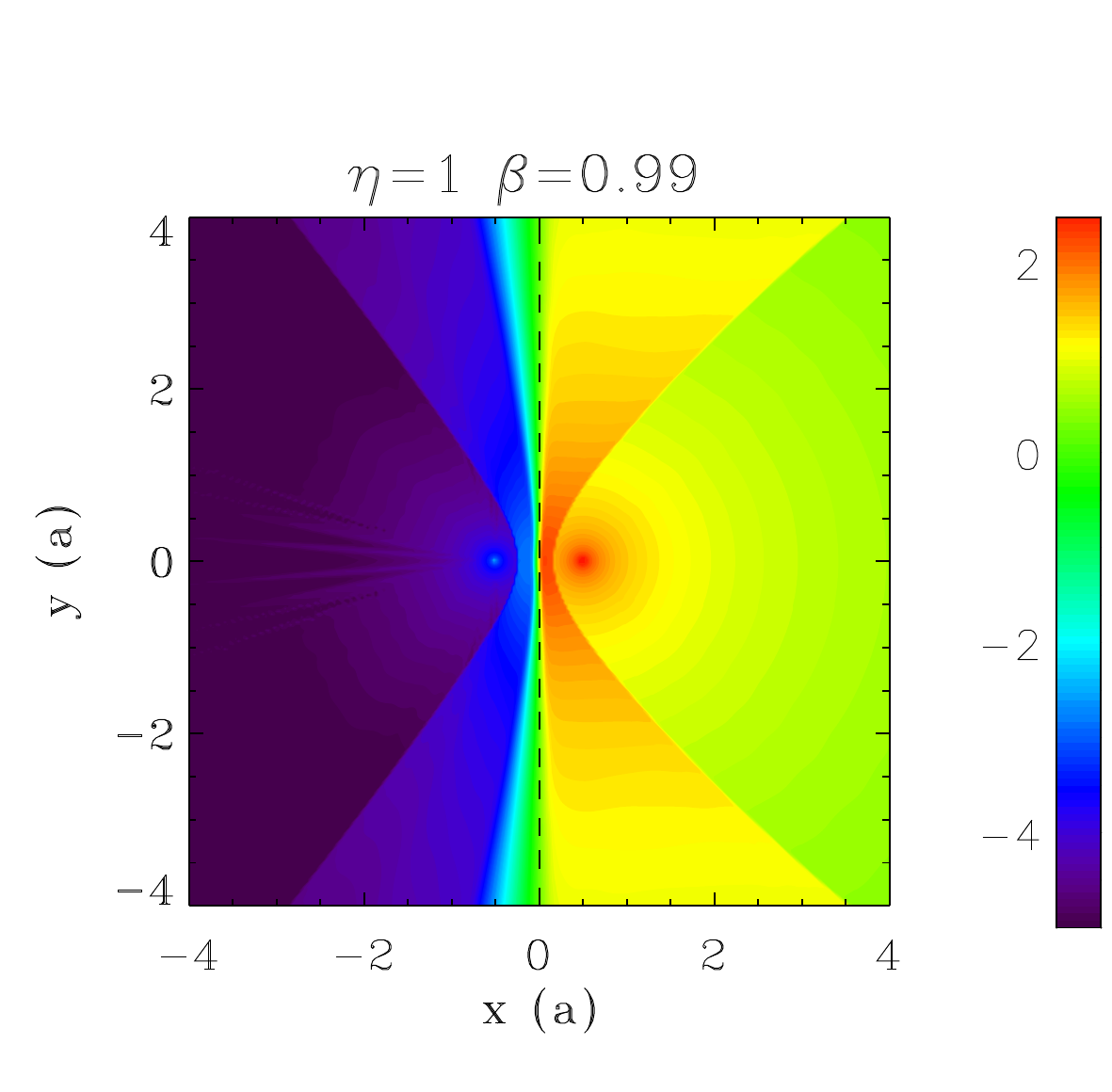}
  \includegraphics[width = .3\textwidth ]{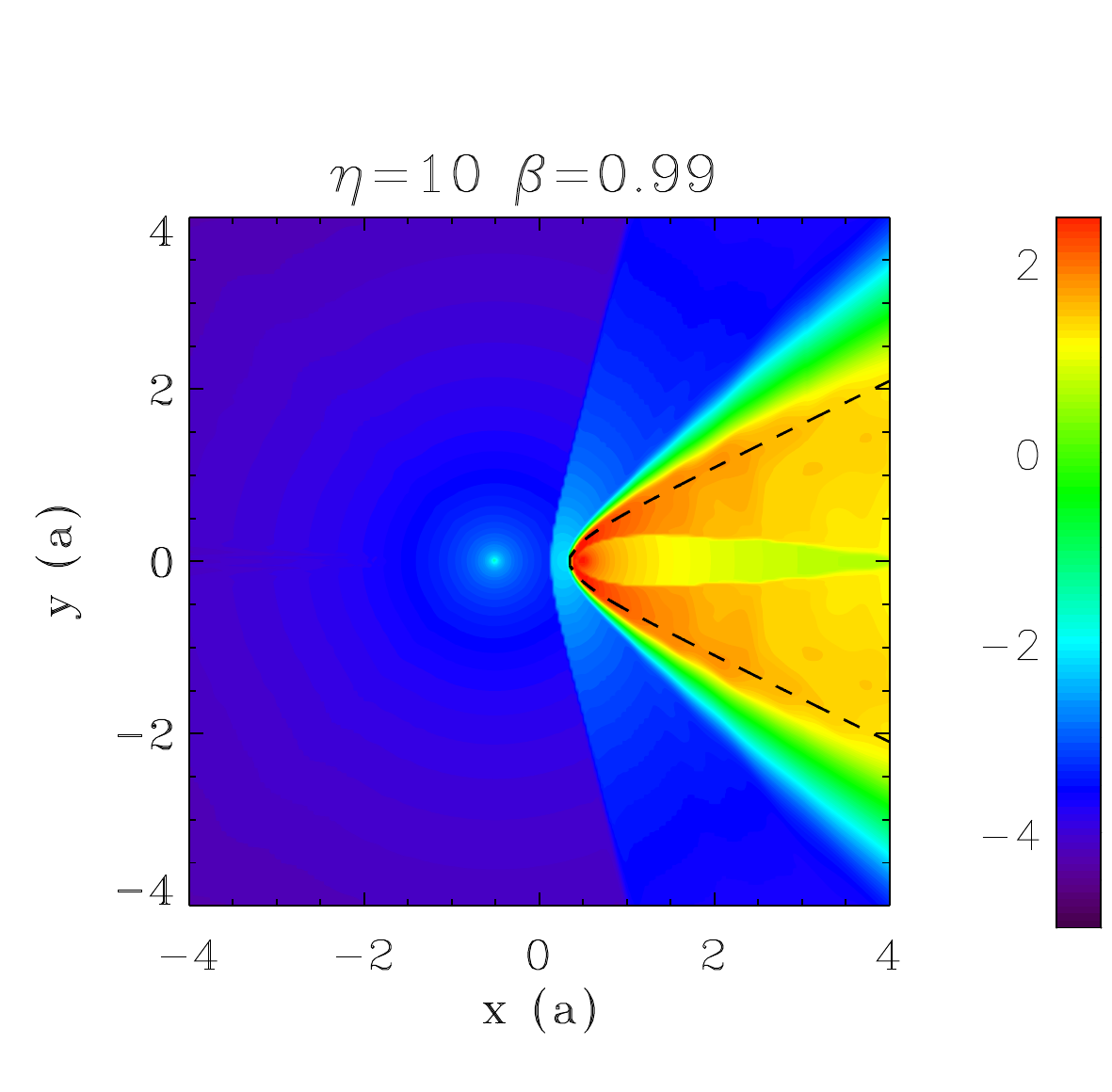}
  \includegraphics[width = .3\textwidth ]{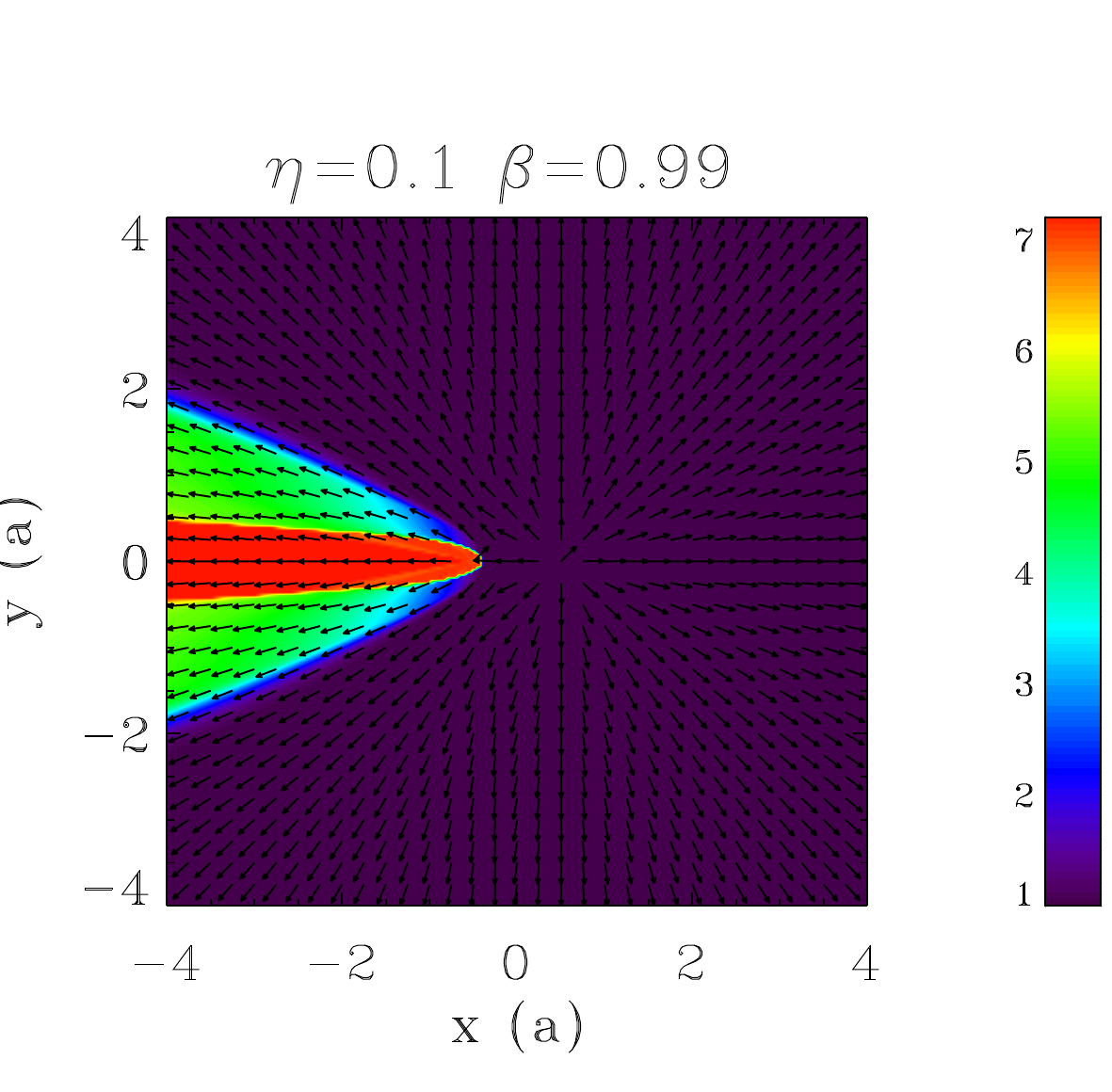}
  \includegraphics[width = .3\textwidth ]{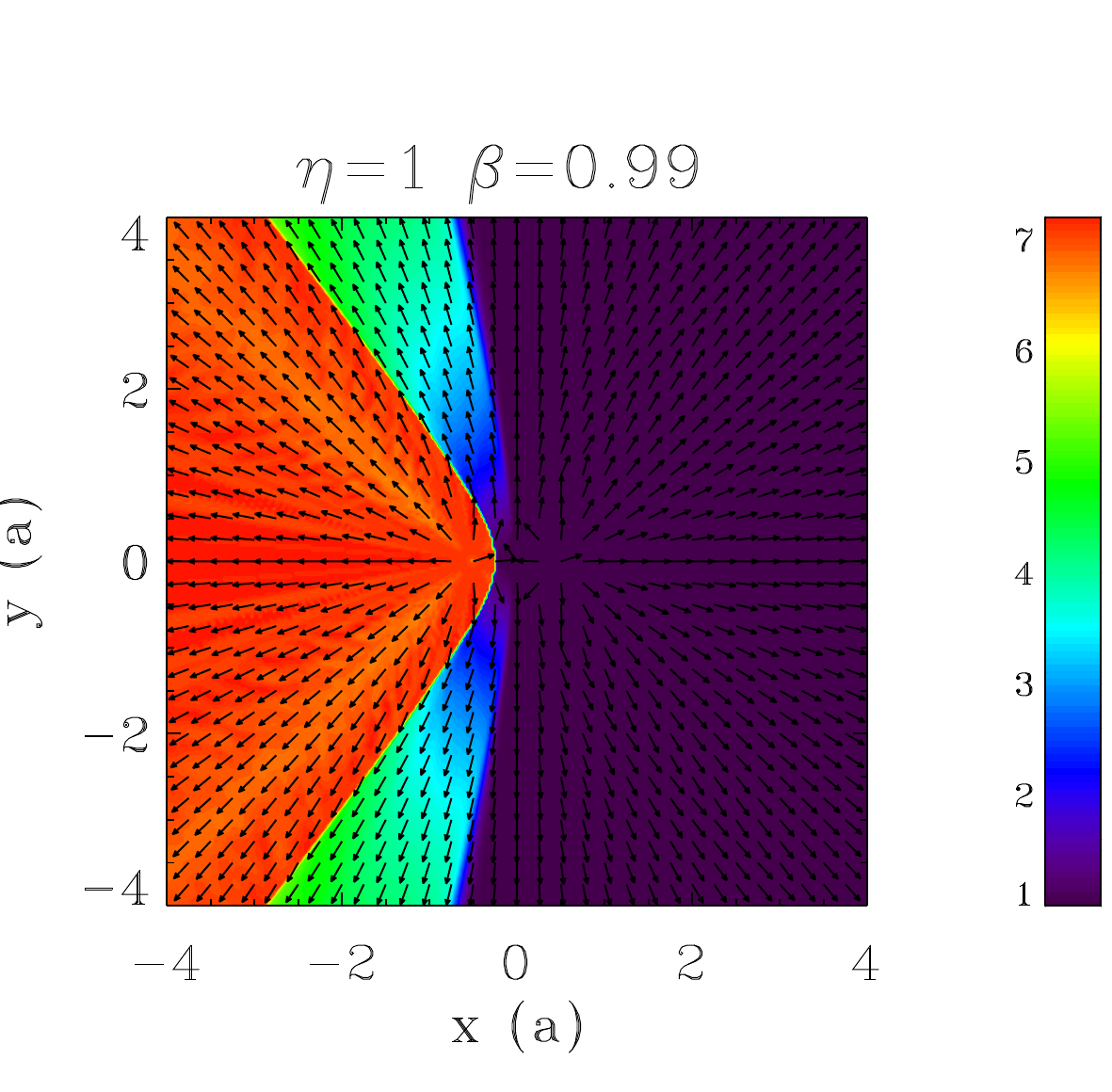}
  \includegraphics[width = .3\textwidth ]{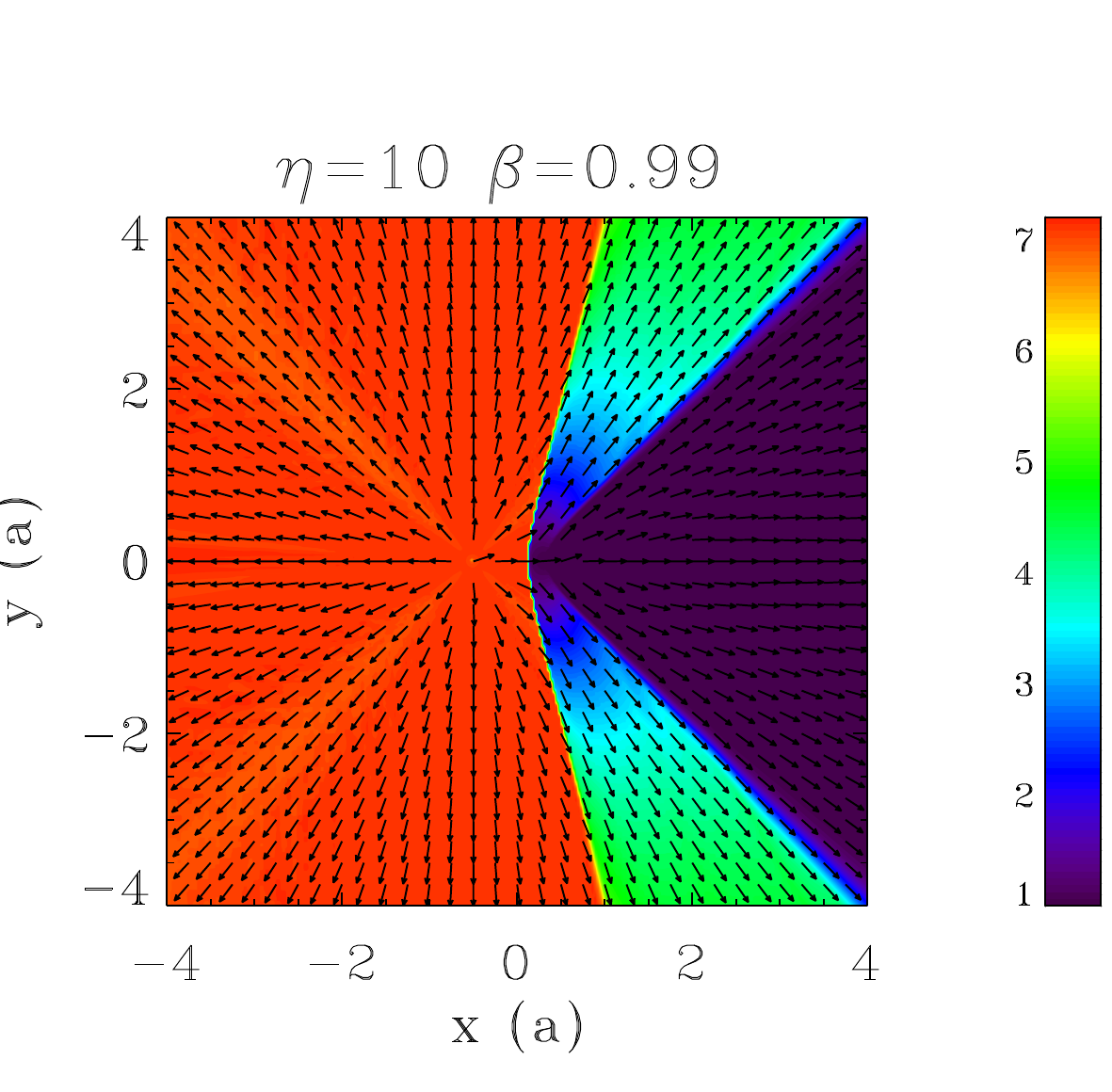}
  \caption{Density (upper row) and Lorentz factor (lower row) maps for simulations with  $\eta_{\rm rel}=0.1,\,1,\,10$ (from left to right)  and $\beta_p=0.99$.  The pulsar is located at $x=-0.5$, the star at $x=0.5$. The density is given in g\,cm$^{-2}$. The arrows show the velocity field.}
  \label{fig:rel_eta}
\end{figure*}

Fig.\,\ref{fig:rel_eta} shows the density map and the Lorentz factor map for  simulations with $\eta_{\rm rel}=0.1,\,1,\,10$ and $\beta_p=0.99$. We set the maximal resolution to $l_{\rm max}=12$ and checked that, for $\eta_{\rm rel}=0.1$ and 10, the positions of the discontinuities does not vary with higher resolution (see \S\ref{num}). The stellar wind is the dense flow on the right, the pulsar wind is located on the left. The dashed lines shows the analytic solution for the contact discontinuity. The solution is  obtained by combining the relativistic definition of the momentum flux ratio $\eta_{rel}$ (Eq. \ref{eq:eta_rel}) with the position of the contact discontinuity derived by \citet{Antokhin:2004hi} for colliding stellar winds. The solution assumes a thin shell geometry where both shocks and contact discontinuity are merged in one single layer. The overall structure appears similar to what is found for non-relativistic flows (see {\em e.g} Fig.\,5 in PaperI). However, relativistic effects have a non-negligible influence on the structure. For instance, the symetry is broken between the cases $\eta_{\rm rel}=0.1$ and $\eta_{\rm rel}=10$, the latter showing a reconfinement shock while the former does not.  The direction of the velocity in the winds is indicated by arrows on the map of the Lorentz factor.  Along the line-of-centres, the winds collide head on, when getting further from the binary the velocity is mostly parallel to the direction of the shocks. This corresponds to the shock tube test simulation (\S\ref{sec:1D_sod}), which required a very high resolution because of the high velocity perpendicular to the  shock normal ($v _{\perp}=0.99$).  The importance of the velocity transverse to the shock normal  is the reason why high resolution is necessary  in simulations with increasing Lorentz factors in the pulsar wind.

\begin{figure}
  \centering
  \includegraphics[width = .4\textwidth ]{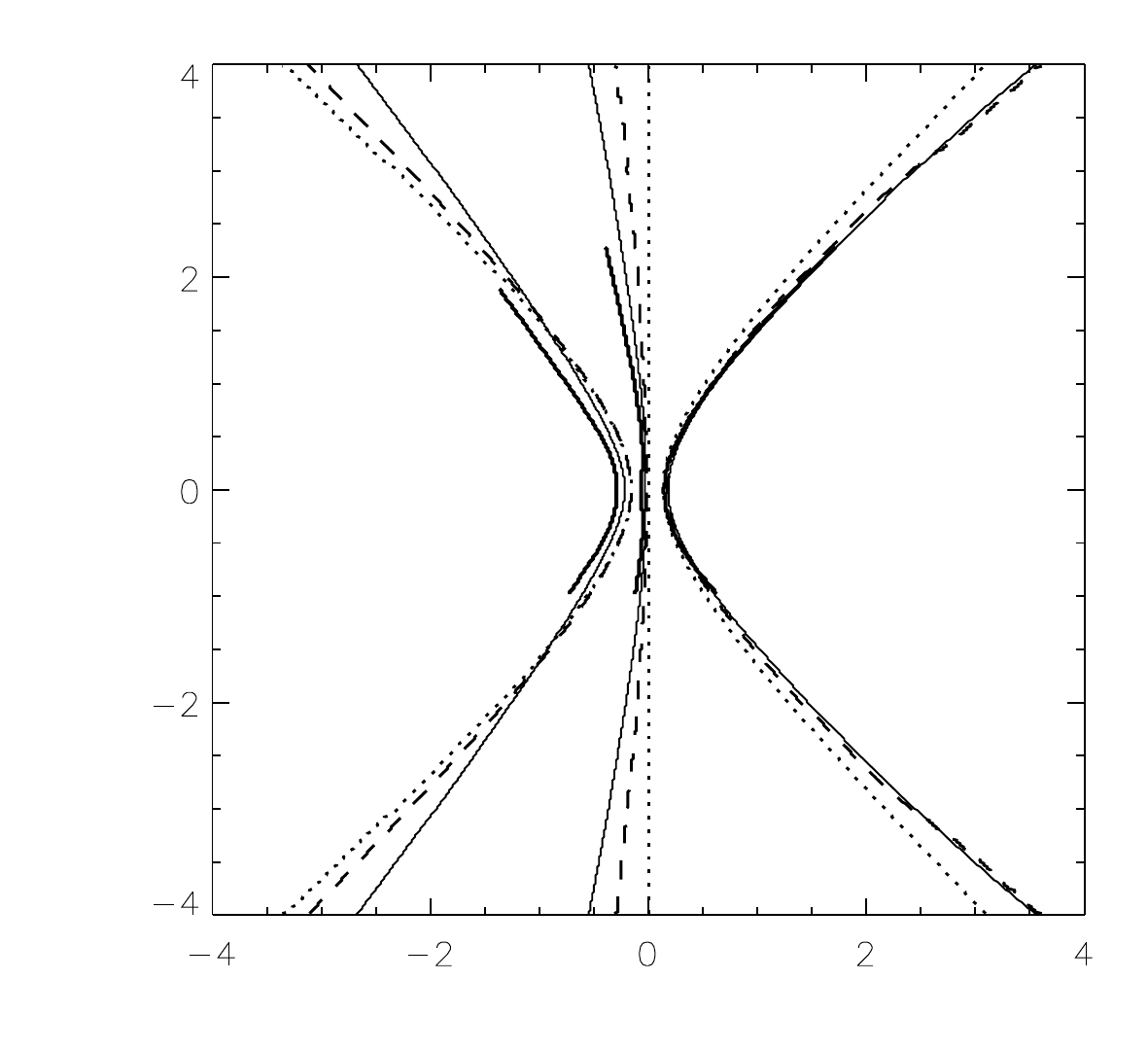}
 \caption{Position of both shocks and the contact discontinuity in a simulation with $\eta_{\rm rel}=1$, in simulations with different values for the velocity of the pulsar wind : $\beta_p$= 0.01 (dotted line),  0.5 (dashed line), 0.99 (solid line) and 0.998 (thick solid line).  For $\beta=0.998$ we used a smaller computational box.}
  \label{fig:rel_comp_lor}
\end{figure}

To quantify the impact of relativistic effects, we carried out simulations with $\eta_{\rm rel}=1$ for increasing values of the pulsar wind velocity $\beta_p=\{0.01,0.5,0.99,0.998\}$. The corresponding Lorentz factors are $\Gamma=\{1.00005,1.15,7.08,\,15.8\}$. Fig.\,\ref{fig:rel_comp_lor} shows the positions of both shocks and the contact discontinuity in the different simulations.   This plot shows that, on the line-of-centres, the location of the contact discontinuity is identical in all simulations and is set by $\eta_{\rm rel}$. At the edges of the box, the contact discontinuity is not exactly midplane between the winds and is closer to the pulsar. We checked that,  for Mach numbers above 10, the contact discontinuity stays midplane in non-relativistic simulations with $\eta=1$ and a different Mach number for each wind.  We interpret the tilt of the CD as a the impact of thermal pressure in the winds.  For a given momentum flux ratio, when the pulsar wind speed is increased, its pressure is decreased, as we keep a fixed Mach number, while  the stellar wind is unchanged. Thus, the higher the Lorentz factor of the pulsar wind, the more the thermal pressure from the stellar wind dominates over the pulsar wind pressure. The tilt of the CD is related to our definition of the pressure in the pulsar wind, which decreases with the Lorentz factor.
 When we set the pressure in the pulsar wind equal to the pressure in the stellar wind, both still being small compared to the ram pressure on the line-of-centres, we find that the CD is exactly on the axis of symmetry.  We consider this case to be unrealistic because the pulsar wind pressure is expected to be negligible due to adiabatic expansion\citep{2009ASSL..357..421K}. The shape of the  shock in the pulsar wind changes with its Lorentz factor, especially far from the binary axis.  In these zones, the velocity is mostly transverse to the shock normal. Contrary to the non-relativistic case, the transverse and normal velocities are tied in the jump conditions via the Lorentz factor and this has an impact on the shock location \citep{PhysRev.74.328}. The position of the  shock in the stellar wind remains unaffected by the speed of the pulsar wind. We conclude that the shape of the interaction region is  affected by relativistic effects, the reason being that the pressure and transverse speed influence the relativistic jump conditions through $h$ and $\Gamma$.  The numerical tests in Appendix B.2 highligh that a very high resolution is needed to properly model the shock propagation speed, when there is a high velocity perpendicular to the shock propagation direction. Although we carefully determined an adequate resolution for our simulations, a numerical effect proper to relativistic simulations cannot be excluded.

\begin{figure}
  \centering
  \includegraphics[width = .4\textwidth ]{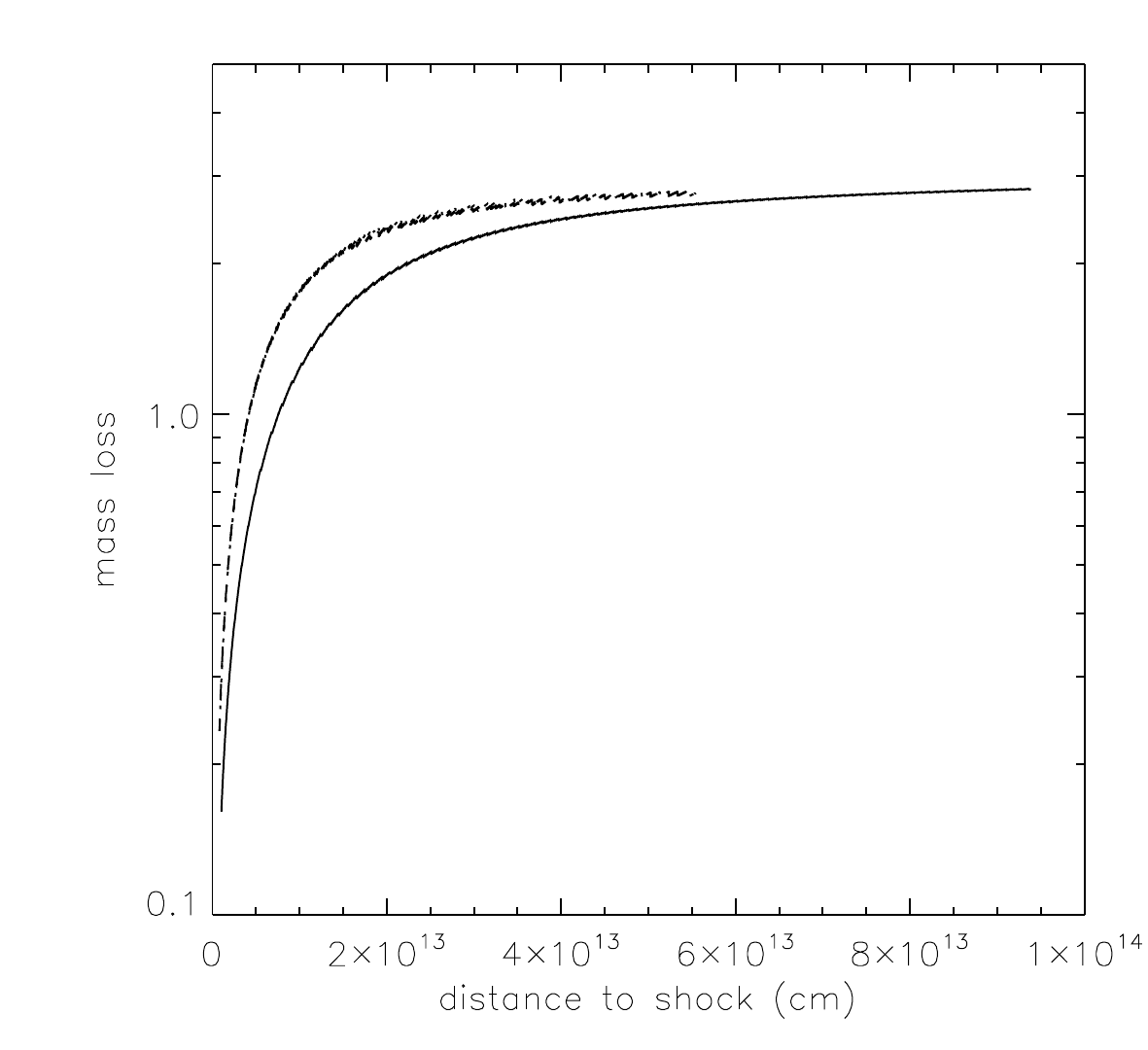}
 \caption{Mass flow $(\dot{M}_p=2\pi s v_p\Gamma_p \rho_p)$ along a streamline in the shocked pulsar wind for $\beta_p=0.01$ (dotted line), 0.5 (dashed line) and 0.99 (solid line). The mass flow is normalized to the initial mass loss rate of the pulsar, we use $\eta_{rel}=1$. The simulations with $\beta_p=0.01$ and $0.5$ give the same results.}
  \label{fig:streamlines}
\end{figure}

To evaluate the impact of the Lorentz factor on the hydrodynamics of the shocked region, we followed the hydrodynamical quantities along streamlines in the shocked pulsar wind. We find that, for all the values of $\beta_p$, starting from the shock, the flow reaccelerates up to its initial velocity. The acceleration tends to a spherical, adiabatic expansion, with $2\pi s v_p \Gamma_p\equiv k\dot{M}_p\simeq$ constant (our simulations are 2D), where $s$ is the curvilinear distance to the shock. Fig.\,\ref{fig:streamlines}  shows the value of $k$ for different speeds of the pulsar wind, for $\eta_{rel}=1$. In all cases the shocked flow reaches $k\approx 3$ on a scale comparable to the binary separation (increasing with $\beta_p$). For $\eta_{\rm rel}=0.1$, we find the same re-acceleration, although with a higher value of $k\simeq 8$. The higher corresponding mass flow is due to the smaller opening angle of the shocked pulsar wind.

\subsection{The Kelvin-Helmholtz instability in $\gamma$-ray binaries}
The stellar wind has a velocity of 3000\,km\,s$^{-1}$ while the pulsar wind is almost two orders of magnitude faster. Similarly to what happens for classical flows, at the contact discontinuity between the winds, the KHI is likely to modify the structure of the flow \citep{1976MNRAS.176..443B,1976MNRAS.176..421T}. \citet{2004PhRvE..70c6304B} found  analytic solutions to the dispersion relation in RHD and showed that, in the frame of the laboratory, the stability criteria are the same as in the classical case, provided one uses the relativistic definition of the Mach number ($\mathcal{M}_{\rm rel}=\mathcal{M}\Gamma/\Gamma_{c_s}$, where $\Gamma_{c_s}$ is the Lorentz factor of the sound speed). In 2D simulations, we thus expect the interface between the winds to be unstable provided $\mathcal{M}_{\rm rel}> \sqrt 2$. 

To verify the impact of the KHI on $\gamma$-ray binaries, we performed a test simulation with $\eta_{\rm rel}=1$ and  $\beta_p=0.99$. \citet{2012arXiv1206.6502R} highlight the importance of an adapted Riemann solver and a low-diffusivity scheme in the study of the KHI. We use the HLLC Riemann solver to enable the development of the KHI at the contact discontinuity between the winds.  This simulation displays a velocity difference $(v_p/v_s\simeq 100)$.  In PaperII, we found that in the classical limit, such a velocity difference leads to important mixing between the wind and destroys the expected large scale spiral structure.

\begin{figure}
  \centering
  \includegraphics[width = .3\textwidth ]{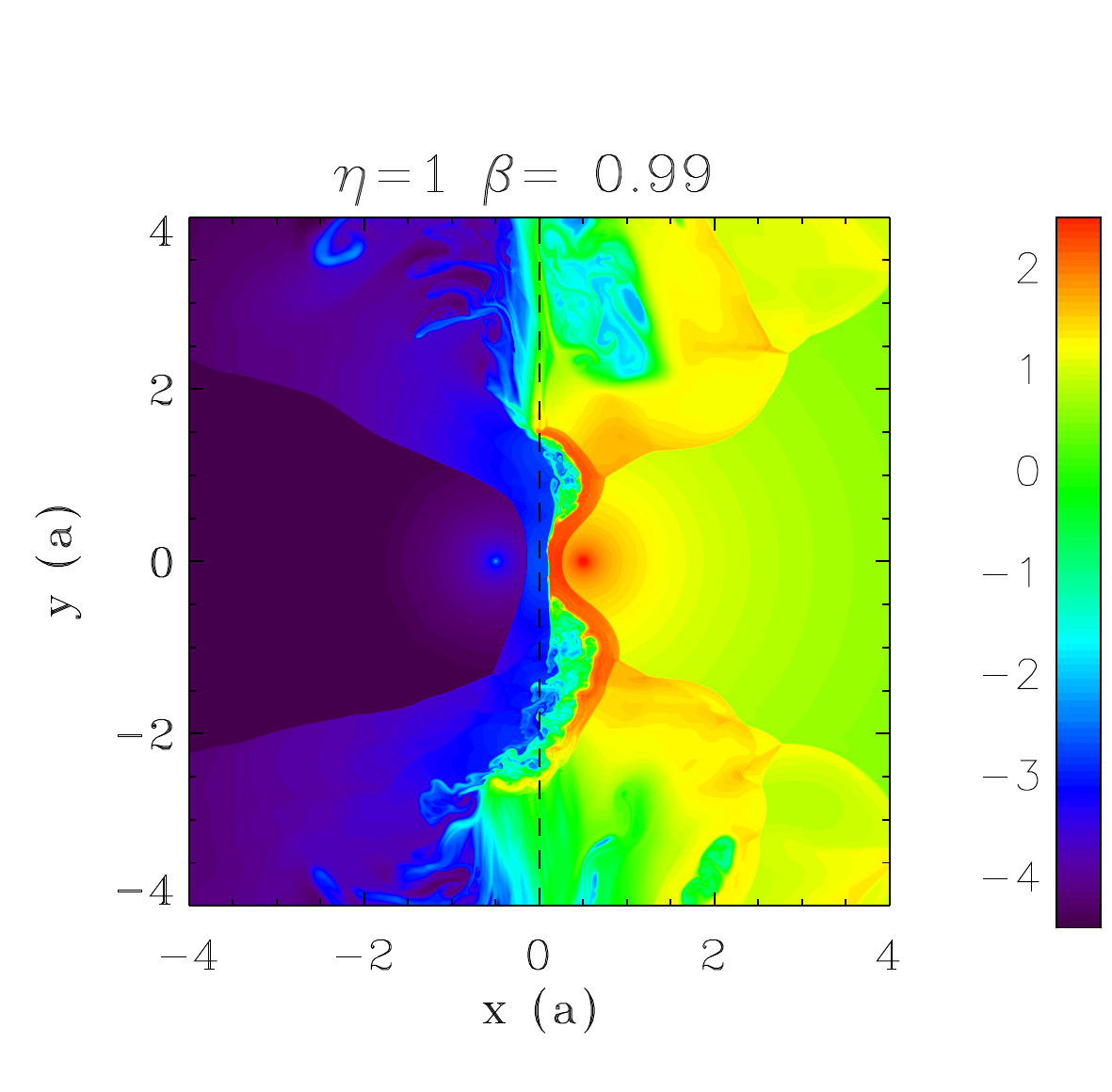}
  \includegraphics[width = .3\textwidth ]{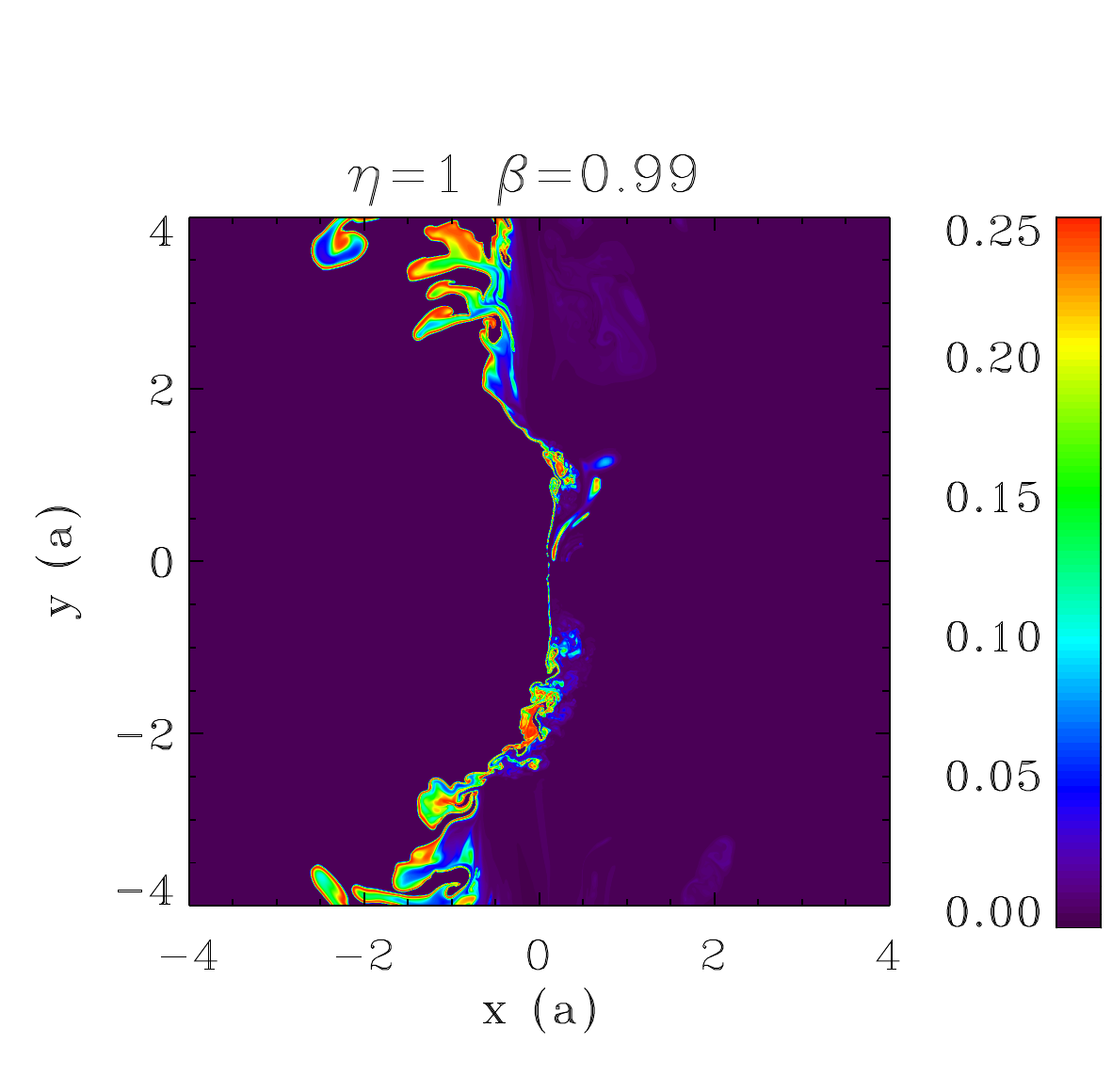}
 \caption{Density (upper panel) and mixing (lower panel) maps for a simulations with $\eta_{\rm rel}=1$ and  $\beta_p=0.99$. }
  \label{fig:rel_KH}
\end{figure}

Fig.~\ref{fig:rel_KH} shows the density map and mixing between the stellar wind and the pulsar wind. The Kelvin-Helmholtz instability is clearly present in the whole colliding wind region.   This is consistent with Fig.7 by \citet{2012A&A...544A..59B}. In their simulation, $\Gamma_p$=10, $\eta_{rel}=.3$ and the stellar wind parameters are very close to ours, making comparison possible. Although orbital motion is included in their work, it has no impact at the location, close to the star,  we are considering.  At a distance of about twice the binary separation, due to the re-acceleration of the pulsar wind, the conditions in the shocked flow are incompatible with the analytic criterion for the development of the KHI. This suggests that  the eddies present far away from the binary are due to the advection of eddies formed closer in. The eddies affect the position of the contact discontinuity but also modify the positions of both shocks. We find important differences in the position of the shock in the pulsar wind between two successive outputs, which indicates variation on a timescale of at most 10 days. The KHI induces some mixing between the winds, although it occurs mostly in low density parts of the shocked region. The overall structure is similar to what we found for colliding non-relativistic stellar winds, for $\eta=1,v_1/v_2=20$ (see Fig. 5 in PaperI) and in a test simulation  with $\eta=1,v_1/v_2=100$ (with the same resolution than the relativistic simulation). Still in the classical limit, the mixed region covers a larger domain, with parts of the faster wind crossing the contact discontinuity and being totally embedded in the slower, denser wind. The mixed region seems smoother, less broken up in the classical limit. Comparison between both simulations is not straightforward as the density jump between both wind is about two orders of magnitude higher in the relativistic simulation, which probably decreases the growth of the intsability.

 \citet{2012A&A...544A..59B} simulated the large scale structure of $\gamma$-ray binaries including orbital motion in 2D. Their simulations of the short period binary LS 5039 show strong instabilities far from the binary, which they partly attribute to the accelerated  transonic flow being deflected by the slower stellar wind, and producing an unstable shocked structure. Whether this effect holds for larger values of the Lorentz factor and  systems with longer orbital periods (such as PSR B1259-63)  is worth investigating further.

\section{Discussion: scaling to realistic pulsar winds}\label{sec:discussion}

The simulations we performed have highlighted some similarities and differences between the interaction of classical stellar winds and the interaction between a stellar wind and a relativistic pulsar wind.  This raises two questions :
\begin{itemize}
\item To what extent can simulations in the classical limit provide information on the structure of gamma-ray binaries? 
\item How can the results of relativistic simulations be extrapolated to more realistic values for the pulsar wind velocity ($\Gamma_p\simeq 10^{4-6}$) ?
\end{itemize}
Pulsar winds are characterized by the spindown power $\dot{E}_p$ carried by the wind, which can be estimated from timing of the pulse period. The corresponding momentum flux ratio can be expressed as a function of $\dot{E}_{p}$ by remarking that $\dot{E}_p=\Gamma \dot{M_p}c^2$  and $v_{p}\approx c$ so 
\begin{equation}
 \label{eq:eta_rel_E}
  \eta_{\rm rel}=\frac{\dot{E}_p}{\dot{M}_s v_s c}\frac{ v_{p}}{c}\approx \frac{\dot{E}_p}{\dot{M}_s v_s c}.
\end{equation}
Using this $\eta_{\rm rel}$ provides a  good indication on the position of the contact discontinuity between the winds and may be used to rescale simulations of non-relativistic colliding stellar winds. Yet, we found that the pulsar wind is more collimated than expected when we carried out relativistic simulations where the pulsar wind pressure $P_p$ is smaller than the stellar wind pressure $P_s$, even though both pressures are negligible compared to the ram pressure term $\rho \Gamma^2 v^2$  on the line-of-centers  (Eq.~\ref{eq:mom_eq_rel}). The difference probably occurs because the relativistic jump conditions are coupled by $\Gamma$ and $h$, with the strongest impact in the shock wings where the speed is mostly transverse to the shock. In contrast, \citet{2008MNRAS.387...63B} found that the location of the contact discontinuity in their relativistic hydrodynamical simulations is identical to the classical case, with $\eta$ calculated using Eq.~\ref{eq:eta_rel_E}. We suspect that they obtain this result because they set the pressure exactly to zero in both winds. Note that we recover the classical result if we set the pressure to the same value in both winds.

Since a realistic pulsar wind has $P_p\approx 0 \ll P_s \neq0 $, our results suggest that pulsar winds are more collimated than assumed by extrapolating from classical results. This may  have a significant impact on the interpretation of the radio VLBI maps of gamma-ray binaries, where the collimated radio emission has been modeled as synchrotron emission from the shocked pulsar wind \citep{2006smqw.confE..52D,2011ApJ...732L..10M,2011A&A...533L...7M,2012A&A...548A.103M}. In particular, \citet{2007A&A...474...15R} noted that the collimation angles in the binary LSI +61\degr303 were too narrow compared to the ones expected in the classical limit. Simulations at higher values of the Lorentz factor are highly desirable to confirm this and fully explore the impact of the wind pressure on shock location.

 To easy comparison with colliding stellar winds, we assumed a constant adiabatic index $\gamma=5/3$ in our simulation. While this is realistic for the stellar wind and (cold) unshocked pulsar wind, the adiabatic index of the (warm) shocked pulsar wind is probably close to 4/3. Performing a test simulation with $\eta_{rel}=1$, $\beta_p=.99$ and $\gamma=4/3$ in both winds, we find that the shocked region in the pulsar wind is about 60 $\%$ denser than when $\gamma=5/3$, and slightly narrower.  These results are  consistent with what we observed in Fig.~\ref{fig:ad_index}.  However, the exact impact on gamma-ray binaries is not straightforward,  as the adiabatic index transitions from the non-relativistic to the ultrarelativistic limit. As can be seen in shock tests in Fig. 5-7 of \citet{2006ApJS..166..410R}, when a relativistic equation of state is used, although the density jump is confined by the two limiting cases, the exact hydrodynamical structure is more complex than a simple intermediate between the case $\gamma=4/3$ and $5/3$.  

Simulations with moderate values for the Lorentz factors provide information on the density and velocity structure of the shocked flow which is useful to calculate the emission properties of the system. We found that the shocked pulsar wind reaccelerates to its initial Lorentz factor and reaches an asymptotic regime with $4 \pi s^2 \rho^*_p v^*_p  \Gamma^{*}_p=k\dot{M_p}=$ constant (or equivalently for our cylindrical 2D geometry $2 \pi s \rho^*_p v^*_p  \Gamma^*_p=k\dot{M_p}=$ constant, with stars for quantities in the shocked region). We find that $k$ is of order unity and is related to the opening angle of the shocked region and seems independent of the Lorentz factor of the pre-shock flow. As the shocked wind accelerates up to its initial pre-shock velocity on a distance of order of a few times the binary separation, we have $v^*_p=v_p$, and this implies that, asymptotically, the density in the shocked wind is given by
\begin{equation}
  \label{eq:rho_shocked2}
  \rho^{*}\simeq\frac{\dot{E}_{p}}{4\pi s^2 c}\frac{k}{v_p^2\Gamma_{p}^2}.
\end{equation}
where we have used $\dot{E}_{p}/c\equiv \Gamma_{p}\dot{M}_{p}v_{p}$.

We have verified this relation by measuring the density ratio  in the shocked pulsar wind for the simulation with $\beta_p=0.5$ and $\beta_p=0.99$. Following Eq.\,\ref{eq:rho_shocked2}, we expect $\rho_{0.5}/\rho_{0.99}\simeq 150$.  At the edge of our simulation domain, where $s\simeq 4 a$, we find that $\rho_{0.5}/\rho_{0.99}$ varies between 140 and 155. This method assumes the flow has reached its terminal velocity. It can be used to estimate the density in the shocked pulsar wind for any value of its Lorentz factor, even using simulations in the non-relativistic limit. However, it is unable to provide precise indications for the properties of the flow close to the binary.  Similar caveats were found by \citet{1996ASPC..100..173K}, who proposed scaling relations to model non-relativistic equivalents of relativistic jets using the mass, energy or momentum flux.  Equating the momentum flux leads to the most satisfactory results on the global morphology of the jet but still underestimates the size of the region with the highest Lorentz factors \citep{1999ApJ...516..729R}. 

The above ideal picture for the shocked flow is likely to be affected by the development of instabilities.  Analytic work and simulations indicate highly relativistic flows are less subject to the Kelvin-Helmholtz instability. \citet{1997ApJ...479..151M,2004A&A...427..431P,2004A&A...427..415P} indicate  that cold jets with high Lorentz factors are more stable than slow, warm jets. In fast flows the timescale for the growth of the KHI is too long with respect to the dynamical time  while in cold flows the sound speed is low and perturbations propagate too slowly. In those particular systems, the specific axisymmetric geometry  plays  a role in the development of particular modes of the instability and makes it hard to directly remap the results to gamma-ray binaries.

Our simulations show that  the KHI  affects the positions of the shocks and induces some mixing between the winds. The most favorable conditions for the development of the KHI occur close to the binary, where the shocked pulsar wind is subsonic and only mildly relativistic. As the downstream velocity cannot be higher than $c/3$ for ultrarelativistic flows,  the conditions for the development of the KHI are likely to be met in real systems (having $\Gamma_p\simeq 10^{4-6}$), at least close to the binary.  The presence of the KHI further may be due to advection only.  Simulations at higher Lorentz factors are necessary to determine whether the size of the unstable region is large enough to enable important growth before the advection phase begins. This is a  key aspect as strong instabilities will induce mixing between the stellar and pulsar wind, which is likely to thermalize the high-energy particles of the pulsar wind and thus decrease their high-energy emission.

\section{Conclusion and Perspectives}\label{sec:conclusion}
We have developed a special relativistic extension to the adaptive-mesh refinement code RAMSES. We have implemented two different second order schemes and Riemann solvers.  So far, only the so-called classical equation of state has been implemented but the code is written in order to make the implementation of more complex equations of state straightforward. The code is fully three-dimensional and allows the use of adaptive mesh refinement. To model high velocity flows, we included the possibility to refine cells according to the gradient of the Lorentz factor.  Our code passes all of the common numerical tests in 1D, 2D and 3D, including the most stringent  tests with high Lorentz factors and velocities perpendicular to the flow.

Using this new relativistic code, we performed a 2D study of the geometry of the $\gamma$-ray binaries, with Lorentz factors up to 16. The relativistic nature of the pulsar wind changes the momentum balance between the winds.  The position of the shock in the pulsar wind is affected  by the impact of the pre-shock wind pressure and transverse velocity on the shock. We find this makes the shocked pulsar wind more collimated than expected using classical hydrodynamics. This relativistic effect is strongest where the flow is essentially along the shock, far from the stagnation point. Comparing simulations with increasing values for the Lorentz factor of the pulsar wind, we find that the shocked pulsar wind re-accelerates up to its initial velocity on a scale of a few times the binary separation. We determine a prescription for the density in the shocked pulsar wind, which can be used for any value of the Lorentz factor. We find that the Kelvin-Helmholtz instability develops at the contact discontinuity affects the structure of the shocked region.  Still, the exact understanding of the instability at realistic values of the pulsar wind Lorentz factor, in the particular geometry of $\gamma$-ray binaries is still to be confirmed. Our simulations provide clues do the understanding of the physics of  gamma-ray binaries. However, detailed comparison with observations requires the inclusion of more complex physical effects.
 
This code will be part of the next public release of RAMSES. It is well suited for the study of many astrophysical flows including gamma-ray bursts, pulsar wind nebulae and jets in galactic binaries or active galactic nuclei.  As an example, the Lorentz factor inferred from the prompt emission of gamma-ray bursts is of order 100, while the afterglow emission is related to a bulk motion of $\Gamma\simeq 10$ (see \citet{2009ARA&A..47..567G} for a review).  Following the fireball model, such systems display a very wide range of lengthscales, from the progenitor to  decelaration radius of the forward shock, including  the reverse shock and the structure of the  internal shocks. Although the fully consistent study of such systems is still out of reach, the possibility for adaptive mesh refinement makes RAMSES a useful tool to model gamma-ray bursts.  Our work on gamma-ray binaries can be easily adapted to the simulation of pulsar winds interacting with the interstellar medium or a supernova remnant. In such cases, the shock occurs further from the pulsar than in gamma-ray binaries, and the typical size of the interaction region is a few  pc. This increase in length scales makes the use of AMR even more fundamental when modelling the structure and instabilities in pulsar wind nebulae.

\appendix
\section{Numerical scheme for RAMSES-RHD}\label{sec:num_details}
This Appendix provides the Jacobian matrices and their eigenstructure for the reconstruction of the primitive variables in the 3D-RHD case. It also details the recovery of the primitive variables.
\subsection{Recovering the primitive variables}\label{sec:app_ctoprim}

To determine the primitive variables $(\rho,v_j,P)^T$, we simplify the method by \citet{2007MNRAS.378.1118M}, initially developed for relativistic magnetohydrodynamics. One rewrites the total energy 
\begin{equation}
  \label{eq:ctoprim}
 E=W'+D-P.  
\end{equation}
and solves an equation on  $W'=W-D=\rho h\Gamma^2-\rho \Gamma$.
 
To avoid numerical problems in the non-relativistic or ultrarelativistic limits, one introduces 
\begin{equation}
  \label{eq:ctoprim2}
 u^2=\frac{M^2}{(W'+D)^2-M^2}=\Gamma^2 v^2  
\end{equation}
 such that the Lorentz factor is given by 
 \begin{equation}
   \label{eq:ctoprim3}
   \Gamma=(1+u^2)^{1/2}.
 \end{equation}
This gives 
\begin{equation}
  \label{eq:ctoprim4}
  W'=\frac{D u^2}{\Gamma+1}+\Gamma^2\frac{\gamma}{\gamma-1}P
\end{equation}
which can be used to replace $P$ in Eq.\,\ref{eq:ctoprim}.
Once $W'$ is found using e.g. a Newton-Raphson method, one recovers the Lorentz factor using Eqs.\,\ref{eq:ctoprim2}-\ref{eq:ctoprim3},  then the density $\rho=D/\Gamma$ and finally pressure using Eq.\,\ref{eq:ctoprim4}. We determine $W'$ with a precision of $10^{-10}$.

 The method can be easily adapted to different equations of states, by simply changing Eq.\,\ref{eq:ctoprim4}. It is  well suited for codes with AMR since an initial guess of the value of $W_0$, which guarantees the positivity of pressure,  can be found analytically \citep{2007MNRAS.378.1118M}. This  avoids storing values of $W'$ between  timesteps, which can be cumbersome on a changing grid. One can determine $W_0$  (which then gives $W_0'=W_0-D$, rewriting the pressure as
\begin{equation}
   \label{eq:pressure}
   P=W-E=\frac{M^2-v^2W^2+4W^2-4EW}{4W}
 \end{equation}
 since $M^2\equiv W^2v^2$.
Pressure is thus guaranteed to be positive when
\begin{equation}
  \label{eq:quadratic}
 M^2-v^2W^2+4W^2-4EW>0 .
\end{equation} 
This happens for any $W>W_+$, with $W_+$ the largest of the two roots of the quadratic equation Eq.\ref{eq:quadratic}. As $v<1$, solving Eq.\,\ref{eq:quadratic} for $v=1$,  implies that the corresponding root $W_{+,1}>W_{+,v}$, which guarantees positive pressure. One thus uses $W_{+,1}$ as the initial guess to start the Newton-Raphson scheme.

Provided the vector of conservative variables is physical, this method converges to a physical primitive state.  The positivity of the density can be verified by $D>$0. Subluminal velocity is guaranteed by  $E^2>M^2+D^2 $. Indeed, using Eq.\,\ref{eq:cons_prim} one has
\begin{equation}
  \label{eq:proof1}
E^2>M^2+D^2\Rightarrow \left(\frac{M}{v}-P\right)^2>M^2+D^2
\end{equation}
This necessarily implies positive pressure and $v<1$. Whenever an non-physical conserved state occurs, we floor the density $\rho=10^{-10}$ and pressure $P=10^{-20}$. This occurs roughly once every $ 10^6$ updates in our gamma-ray binary simulations. It occurs in the unshocked pulsar wind,  where the density and pressure are lowest, and the Lorentz factor the highest. It mostly occurs along the line of centers of the binary, as the Cartesian grid is unable to perfectly model the spherical symmetry of the wind. The number of failures strongly reduces with resolution. \citet{2011ApJS..193....6B} note a few failures when pressure is very low and suggest to use the entropy instead of the energy as conserved variable in those cases.  Reconstruction by using the neighbouring cells is very cumbersome due to the AMR structure in RAMSES, so we have not considered this option.

 \subsection{Computing the timestep}
The timestep is determined by the Courant condition for unsplit schemes

\begin{equation}
  \label{eq:cfl}
  \Delta t=C_{CFL}\times \min \left(\frac{\Delta x}{\lambda_x + \lambda_y + \lambda_z}\right),
\end{equation}

 where $C_{CFL}$ is the Courant number and $\lambda_k$ is the propagation speed along direction $k$ determined by 

\begin{equation}
\lambda_k=\max(|\lambda_+|,|\lambda_-|),
\end{equation}

with
\begin{equation}
  \lambda_{\pm}=\frac{v_{\parallel}(1-c_s^2)+\sqrt{(1-v^2)(1-v_{\parallel}^2-v_{\perp}c_s^2})}{1-v^2c_s^2},\\
\end{equation}
 where $v_{\parallel}$ and $v_{\perp}$ are the velocity parallel and perpendicular to the direction $k$. All the simulations in this paper were performed using $C_{CFL}=.8$. 

 Our determination of the timestep is more restrictive than the more commonly used (see e.g. \citet{2012ApJS..198....7M})
\begin{equation}
  \label{eq:cfl}
  \Delta t=C_{CFL}\times \min \left(\frac{\Delta x}{\lambda_x}, \frac{\Delta x}{\lambda_y} , \frac{\Delta x}{\lambda_z}\right).
\end{equation}
Using the above expression for the timestep, we find that the simulation of the 3D jet (Appendix \ref{sec:jet}) shows no failure for $C_{CFL}\leq .5$, which is comparable to other codes \citep{2006ApJS..164..255Z}.

\subsection{Jacobian matrices of the 3D-RHD equations}\label{sec:app_jac}
This subsection will be available online only.
 One has
\begin{eqnarray}
  \label{eq:evolution}
 \partial \mathbf{q}&=&-\mathbf{A}_x\frac{\partial \mathbf{q}}{\partial x} dt -\mathbf{A}_y \frac{\partial \mathbf{q}}{\partial y} dt -\mathbf{A}_z\frac{\partial \mathbf{q}}{\partial z} dt.\\
\end{eqnarray}
Following \citet{1994A&A...282..304F} for a 1D flow with transverse velocity and the general definition of the enthalpy 
\begin{eqnarray}
  \frac{\partial h}{\partial x} &=&\frac{\partial h}{\partial \rho}\frac{\partial \rho}{\partial x} +\frac{\partial h}{\partial P}\frac{\partial P}{\partial x} \equiv \chi \frac{\partial \rho}{\partial x} +\kappa\frac{\partial P}{\partial x}
\end{eqnarray}
With these definitions, the sound speed can be written as
\begin{eqnarray}
{c^2_s}=\frac{\rho\chi}{h\left(1-\rho\kappa\right)}.
\end{eqnarray}
For the classical equation of state, one has 
\begin{eqnarray}
  \chi&=&-\frac{\gamma}{ \gamma -1}\frac{P}{\rho^2} \\
  \kappa&=&\frac{\gamma}{ \gamma -1}\frac{1}{\rho} .
\end{eqnarray}

The complete Jacobian matrix  is given by 

\begin{equation}
  \mathbf{A}_x=
  \begin{pmatrix}
c_{11}  & c_{12} & c_{13} & c_{14} & c_{15}\\
c_{21}  & c_{22} & c_{23} & c_{24} & c_{25}\\
c_{31}  & c_{32} & c_{33} & c_{34} & c_{35}\\
c_{41}  & c_{42} & c_{43} & c_{44} & c_{45}\\
c_{51}  & c_{52} & c_{53} & c_{54} & c_{55}\\
 \end{pmatrix}
.
\end{equation}

\begin{eqnarray*}
  Nc_{11}&=& v_xN\\
  Nc_{12}&=& \rho h \Gamma^2(\rho\kappa -1)\\
  Nc_{13}&=&0\\
  Nc_{14}&=&0\\
  Nc_{15}&=&v_x(1-\rho\kappa)\\
  Nc_{21}&=&0\\
  Nc_{22}&=&v_x\Gamma^2(\rho h \kappa -h+\rho \chi)\\
  Nc_{23}&=&0\\
  Nc_{24}&=&0\\
  Nc_{25}&=&\frac{1}{\rho h \Gamma^2}(\rho h \Gamma^2\kappa (1-v_x^2) -h -h\Gamma^2(v_y^2+v_z^2)+\rho \chi \Gamma^2(v_y^2+v_z^2))\\
  Nc_{31}&=&0\\
  Nc_{32}&=&v_y \rho \chi\\
  Nc_{33}&=&v_xN\\
  Nc_{34}&=&0\\
  Nc_{35}&=&-\frac{v_x v_y (-h +\rho \chi +\rho h \kappa)}{\rho h}\\
  Nc_{41}&=&0\\
  Nc_{42}&=&v_z \rho \chi \\
  Nc_{43}&=&0 \\
  Nc_{44}&=&v_xN\\
  Nc_{45}&=&-\frac{v_x v_z (-h +\rho \chi +\rho h \kappa)}{\rho h}\\
  Nc_{51}&=&0\\
  Nc_{52}&=&-\rho ^2h \Gamma^2 \chi\\
  Nc_{53}&=&0\\
  Nc_{54}&=&0\\
  Nc_{55}&=&v_x\Gamma^2(\rho \chi -h +h \rho \kappa)
\end{eqnarray*}
with $N=(-h+h\rho\kappa+\rho\chi v^2)\Gamma^2$.

The  matrix $\mathbf{A}_y$  is given by
\begin{eqnarray*}
Nc_{11}&=& v_yN\\
Nc_{12}&=& 0\\
Nc_{13}&=& \rho h \Gamma^2(\rho \kappa-1)\\
Nc_{14}&=& 0\\
Nc_{15}&=&v_y(-\rho\kappa\Gamma^2+\rho\kappa\Gamma^2v^2+1)\\
Nc_{21}&=&0\\
Nc_{22}&=&v_yN\\
Nc_{23}&=&v_x\rho\chi\\
Nc_{24}&=&0\\
Nc_{25}&=&\frac{-(v_xv_y(-h+\rho\chi+h\rho   \kappa))}{\rho h}\\
Nc_{31}&=&0\\
Nc_{32}&=&0\\
Nc_{33}&=&v_y\Gamma^2(\rho \chi -h +h \rho \kappa)\\
Nc_{34}&=&0\\
Nc_{35}&=&\frac{1}{\rho h \Gamma^2}(\rho h \Gamma^2\kappa(1- v_y^2) -h -h\Gamma^2(v_x^2+v_z^2)+\rho \chi \Gamma^2(v_x^2+v_z^2))\\
Nc_{41}&=&0\\
Nc_{42}&=&0\\
Nc_{43}&=&v_z\rho\chi\\
Nc_{44}&=&v_yN\\
Nc_{45}&=&\frac{-(v_xv_z(-h+\rho\chi+\rho h\kappa)}{\rho h}\\
Nc_{51}&=&0\\
Nc_{52}&=&0\\
Nc_{53}&=&-\rho^2 h\Gamma^2\chi\\
Nc_{54}&=&0\\
Nc_{55}&=&v_y\Gamma^2(\rho \chi -h +h \rho \kappa).
\end{eqnarray*}

Along $z$ one has $\mathbf{A}_z$  given by 
\begin{eqnarray*}
Nc_{11}&=& v_zN\\
Nc_{12}&=& 0\\
Nc_{13}&=& 0\\
Nc_{14}&=& h\rho\Gamma^2(\rho\kappa-1)\\
Nc_{15}&=&v_z(-\rho \Gamma^2\kappa+\rho v^2\Gamma^2\kappa+1)\\
Nc_{21}&=&0\\
Nc_{22}&=&v_zN\\
Nc_{23}&=&0\\
Nc_{24}&=&v_z\rho \chi\\
Nc_{25}&=&\frac{-(v_xv_z(-h+\rho\chi+\rho h\kappa)}{\rho h}\\
Nc_{31}&=&0\\
Nc_{32}&=&0\\
Nc_{33}&=&v_zN\\
Nc_{34}&=&v_y\rho\chi\\
Nc_{35}&=&\frac{-(v_yv_z(-h+\rho\chi+\rho h\kappa)}{\rho h}\\
Nc_{41}&=&0\\
Nc_{42}&=&0\\
Nc_{43}&=&0\\
Nc_{44}&=&v_z\Gamma^2(\rho \chi -h +h \rho \kappa)\\
Nc_{45}&=&\frac{1}{\rho h \Gamma^2}(\rho h \Gamma^2\kappa(1- v_z^2) -h -h\Gamma^2(v_x^2+v_y^2)+\rho \chi \Gamma^2(v_x^2+v_y^2))\\
Nc_{51}&=&0\\
Nc_{52}&=&0\\
Nc_{53}&=&0\\
Nc_{54}&=&-\rho^2 h \Gamma^2\chi\\
Nc_{55}&=&v_z\Gamma^2(\rho \chi -h +h \rho \kappa).
\end{eqnarray*}
\subsection{Eigenstructure}

The  eigenvalues of $\mathbf{A}_x$ are (see e.g. \citet{1996MNRAS.278..586F})
\begin{eqnarray}
  \label{eq:eigenvalues}
  \lambda_-&=&v_x\frac{(1-c_s^2)-c_s\Gamma^{-1}\omega}{1-c_s^2v^2}\\
\lambda_{0x}&=&v_x\\
\lambda_{0y}&=&v_x\\
\lambda_{0z}&=&v_x\\
  \lambda_+&=&v_x\frac{(1-c_s^2)+c_s\Gamma^{-1}\omega}{1-c_s^2v^2}.\\
\end{eqnarray}

with 
\begin{equation}
  \label{eq:omega}
  \omega=\sqrt{1-v_x^2-c_s(v_y^2+v_z^2)}
\end{equation}
The left eigenvectors are given by 
\begin{eqnarray}
  \label{eq:left_eigen}
  L_-&=&(0,-\frac{\rho \Gamma}{2c_s\omega}, 0, 0, \frac{1}{2 c_s^2 h})\\
  L_{0x}&=&(1,0,0,0,-\frac{1}{c_s^2h})\\
  L_{0y}&=&(0, \frac{v_xv_y}{1-v_x^2},1,0,\frac{v_y}{(1-v_x^2)\Gamma^2\rho h})\\
  L_{0z}&=&(0, \frac{v_xv_z}{1-v_x^2},0,1,\frac{v_z}{(1-v_x^2)\Gamma^2\rho h})\\
  L_+&=&(0,\frac{\rho \Gamma\omega}{2 c_s}, 0, 0, \frac{1}{2 c_s^2 h}).\\
\end{eqnarray}

The right eigenvectors are given by 
\begin{eqnarray}
  \label{eq:right_eigen}
  R_-&=&(1,-\frac{c_s\omega}{\rho \Gamma},  \frac{-v_y c_s(\Gamma\omega v_x+c_s)  }{\rho \Gamma^2(1-v_x^2)},\frac{-v_z c_s(\Gamma\omega v_x+c_s) }{\rho\Gamma^2(1-v_x^2)}, c_s^2h)^{T}\\
  R_{0x}&=&(1,0,0,0,0)^T\\
  R_{0y}&=&(0,0,1,0,0)^T\\
  R_{0z}&=&(0,0,0,1,0)^T\\
  R_+&=&(1, \frac{c_s\omega}{\rho \Gamma}, \frac{v_y c_s(\Gamma\omega v_x-c_s)  }{\rho \Gamma^2(1-v_x^2)},\frac{v_z c_s(\Gamma\omega v_x-c_s) }{\rho\Gamma^2(1-v_x^2)}, c_s^2h)^{T}\\
\end{eqnarray}

Similarly for $\mathbf{A}_y$ one has

\begin{eqnarray}
  \label{eq:eigenvalues}
  \lambda_-&=&v_y\frac{(1-c_s^2)-c_s\Gamma^{-1}\omega}{1-c_s^2v^2}\\
\lambda_{0x}&=&v_y\\
\lambda_{0y}&=&v_y\\
\lambda_{0z}&=&v_y\\
  \lambda_+&=&v_y\frac{(1-c_s^2)+c_s\Gamma^{-1}\omega}{1-c_s^2v^2}.\\
\end{eqnarray}
with 
\begin{equation}
  \label{eq:omega}
  \omega=\sqrt{1-v_y^2-c_s(v_x^2+v_z^2)}
\end{equation}
The left eigenvectors are given by 
\begin{eqnarray}
  \label{eq:left_eigen}
  L_-&=&(0,0, -\frac{\rho \Gamma}{2 c_s \omega}, 0, \frac{1}{2 c_s^2 h})\\
  L_{0x}&=& (0,1,\frac{v_yv_x}{1-v_y^2},0,\frac{v_x}{(1-v_y^2)\Gamma^2\rho h})   \\
  L_{0y}&=& (1,0,0,0,-\frac{1}{c_s^2h})\\
  L_{0z}&=& (0, 0, \frac{v_yv_z}{1-v_y^2},1,\frac{v_z}{(1-v_y^2)\Gamma^2\rho h})\\
  L_+&=&(0,0,\frac{\rho \Gamma} { 2 c_s \omega}, 0, \frac{1}{2 c_s^2 h}).\\
\end{eqnarray}

The right eigenvectors are given by 
\begin{eqnarray}
  \label{eq:right_eigen}
  R_-&=&(1,\frac{-v_x c_s(\Gamma\omega v_y+c_s)  }{\rho \Gamma^2(1-v_y^2)} ,-\frac{c_s\omega}{\rho \Gamma},\frac{-v_z c_s(\Gamma\omega v_y+c_s)  }{\rho \Gamma^2(1-v_y^2)} , c_s^2h)^{T}\\
  R_{0x}&=&(0,1,0,0,0)^T\\
  R_{0y}&=&(1,0,0,0,0)^T\\
  R_{0z}&=&(0,0,0,1,0)^T\\
  R_+&=&(1,\frac{v_x c_s(\Gamma\omega v_y-c_s)  }{\rho \Gamma^2(1-v_y^2)}  ,+\frac{c_s\omega}{\rho \Gamma},\frac{v_z c_s(\Gamma\omega v_y-c_s)  }{\rho \Gamma^2(1-v_y^2)}   , c_s^2h)^{T},\\
\end{eqnarray}

For $\mathbf{A}_z$ one has

\begin{eqnarray}
  \label{eq:eigenvalues}
  \lambda_-&=&v_z\frac{(1-c_s^2)-c_s\Gamma^{-1}\omega}{1-c_s^2v^2}\\
\lambda_{0x}&=&v_z\\
\lambda_{0y}&=&v_z\\
\lambda_{0z}&=&v_z\\
  \lambda_+&=&v_z\frac{(1-c_s^2)+c_s\Gamma^{-1}\omega}{1-c_s^2v^2}.\\
\end{eqnarray}

with 
\begin{equation}
  \label{eq:omega}
  \omega=\sqrt{1-v_z^2-c_s(v_x^2+v_y^2)}
\end{equation}

The left eigenvectors are given by 
\begin{eqnarray}
  \label{eq:left_eigen}
  L_-&=&(0,0,0, -\frac{\rho \Gamma}{2 c_s \omega}, \frac{1}{2 c_s^2 h})\\
  L_{0x}&=& (0, 1,0, \frac{v_xv_z}{1-v_z^2},\frac{v_x}{(1-v_z^2)\Gamma^2)\rho h})\\
  L_{0y}&=& (0,0, \frac{v_y}{(1-v_z^2)\Gamma^2\rho h},\frac{v_yv_z}{1-v_z^2},0,)\\
  L_{0z}&=& (1, 0,0 ,0 ,-\frac{1}{c_s^2h})\\
  L_+&=&(0,0,0, \frac{\rho \Gamma}{2 c_s \omega}, \frac{1}{2 c_s^2 h}).\\
\end{eqnarray}

The right eigenvectors are given by 
\begin{eqnarray}
  \label{eq:right_eigen}
  R_-&=&(1, \frac{-v_x c_s(\Gamma\omega v_z+c_s)  }{\rho \Gamma^2(1-v_z^2)}    , \frac{-v_y c_s(\Gamma\omega v_z+c_s)  }{\rho \Gamma^2(1-v_z^2)}  ,-\frac{c_s\omega}{\rho \Gamma},  c_s^2h)^{T}\\
  R_{0x}&=&(0,1,0,0,0)^T\\
  R_{0y}&=&(0,0,1,0,0)^T\\
  R_{0z}&=&(1,0,0,0,0)^T\\
  R_+&=&(1,\frac{v_x c_s(\Gamma\omega v_z-c_s)  }{\rho \Gamma^2(1-v_y^2)}  ,\frac{v_x c_s(\Gamma\omega v_z-c_s)  }{\rho \Gamma^2(1-v_y^2)}  , c_s^2h)^{T}.\\
\end{eqnarray}

\section{Numerical tests}\label{sec:tests}
In this section, we present a set of numerical tests that validate our new relativistic code. Unless stated differently, all the simulations were performed with Courant number $C_{CFL}=0.8$, the \textit{minmod} slope limiter in the MUSCL scheme and for the AMR prolongations, and the HLLC Riemann solver. The adiabatic index is set to 5/3. Refinement is based on density and velocity gradients.
 
\subsection{1D Sod test}\label{sec:1D_sod}
One initially starts with two different media separated by an interface located at $x=0.5$. When the simulation begins, the interface is removed and the flow evolves freely following a selfsimilar structure.  The flow decays into three waves, usually a rarefaction, a contact discontinuity and a shock. These tests are very common because easy to implement and comparisons with analytic solutions are possible (see \citet{Marti1994} for pioneering work and \cite{Rezolla2003} for an elegant  version including transverse velocities). 

We reproduce the test of \S\ref{sec:AMR},   with a maximal Lorentz factor of 120. Contrary to the Newtonian case, transverse velocities do impact the shock structure in RHD and reduce the shock velocity.   Several groups have shown that tests with high transverse velocities are very stringent and show satisfactory results only at very high resolution \citep{ 2005MNRAS.364..126M,2006ApJS..164..255Z}. 

\begin{figure*}[h]
  \centering
  \includegraphics[width = .32\textwidth ]{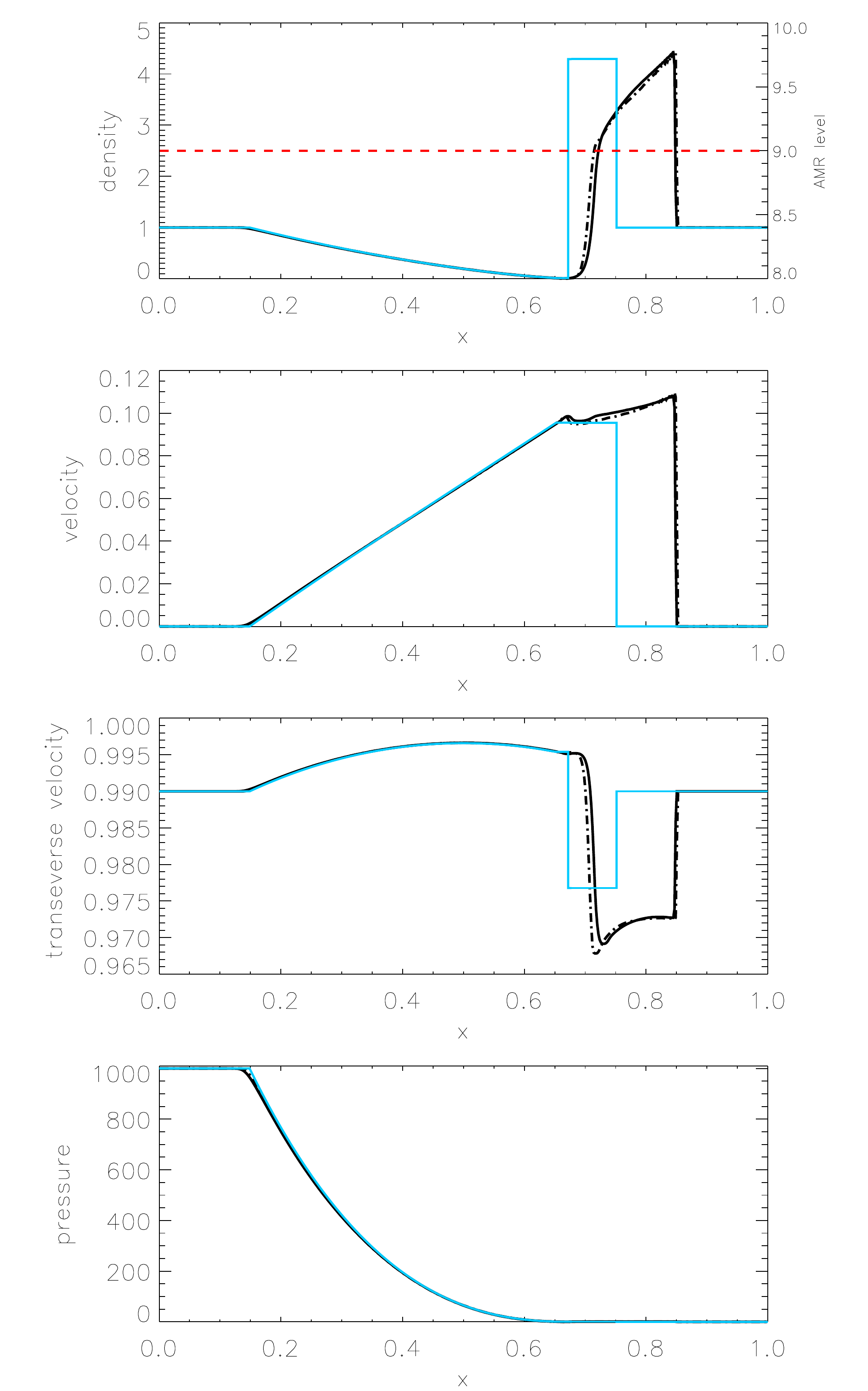}
  \includegraphics[width = .32\textwidth ]{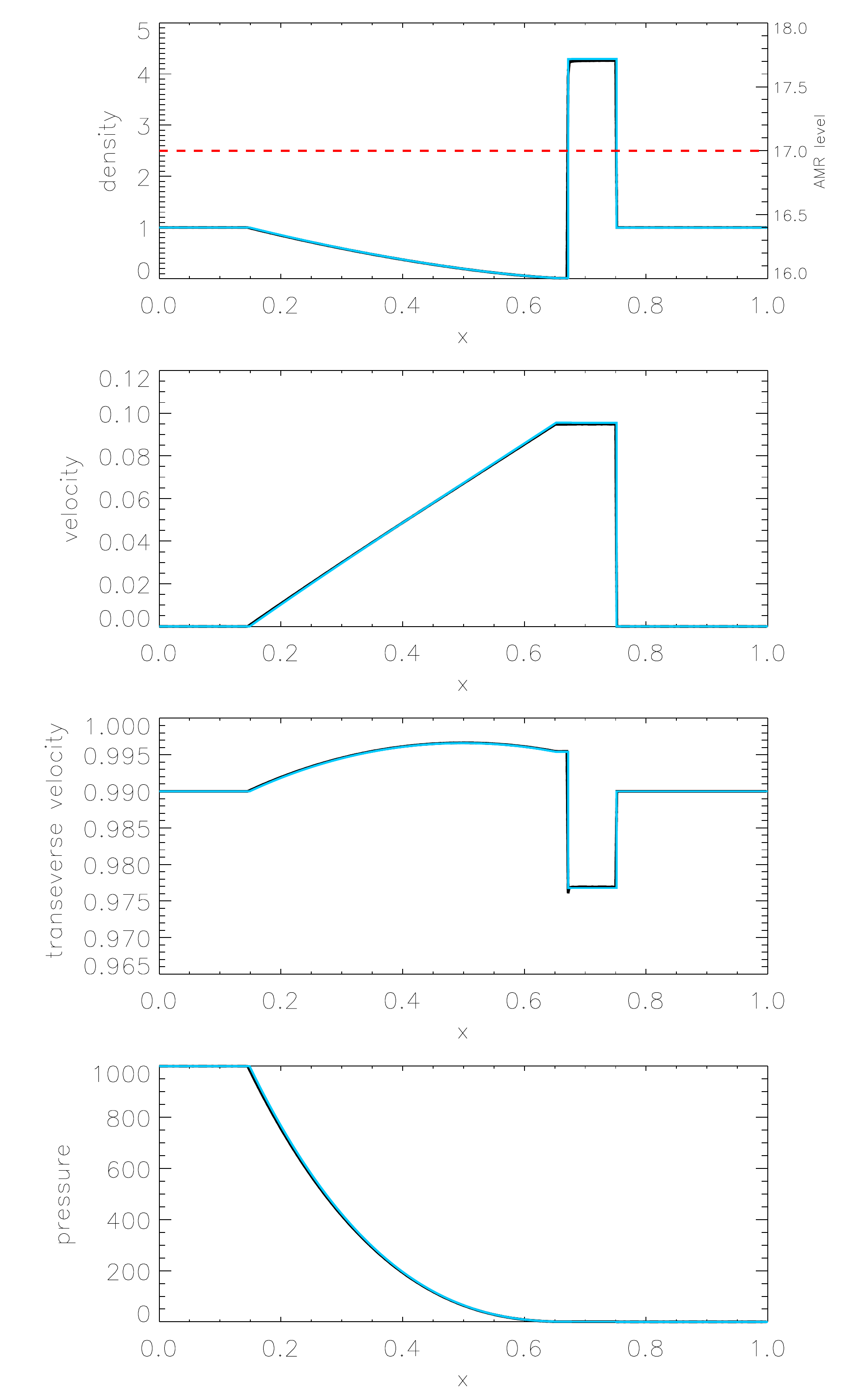}  
  \includegraphics[width = .32\textwidth ]{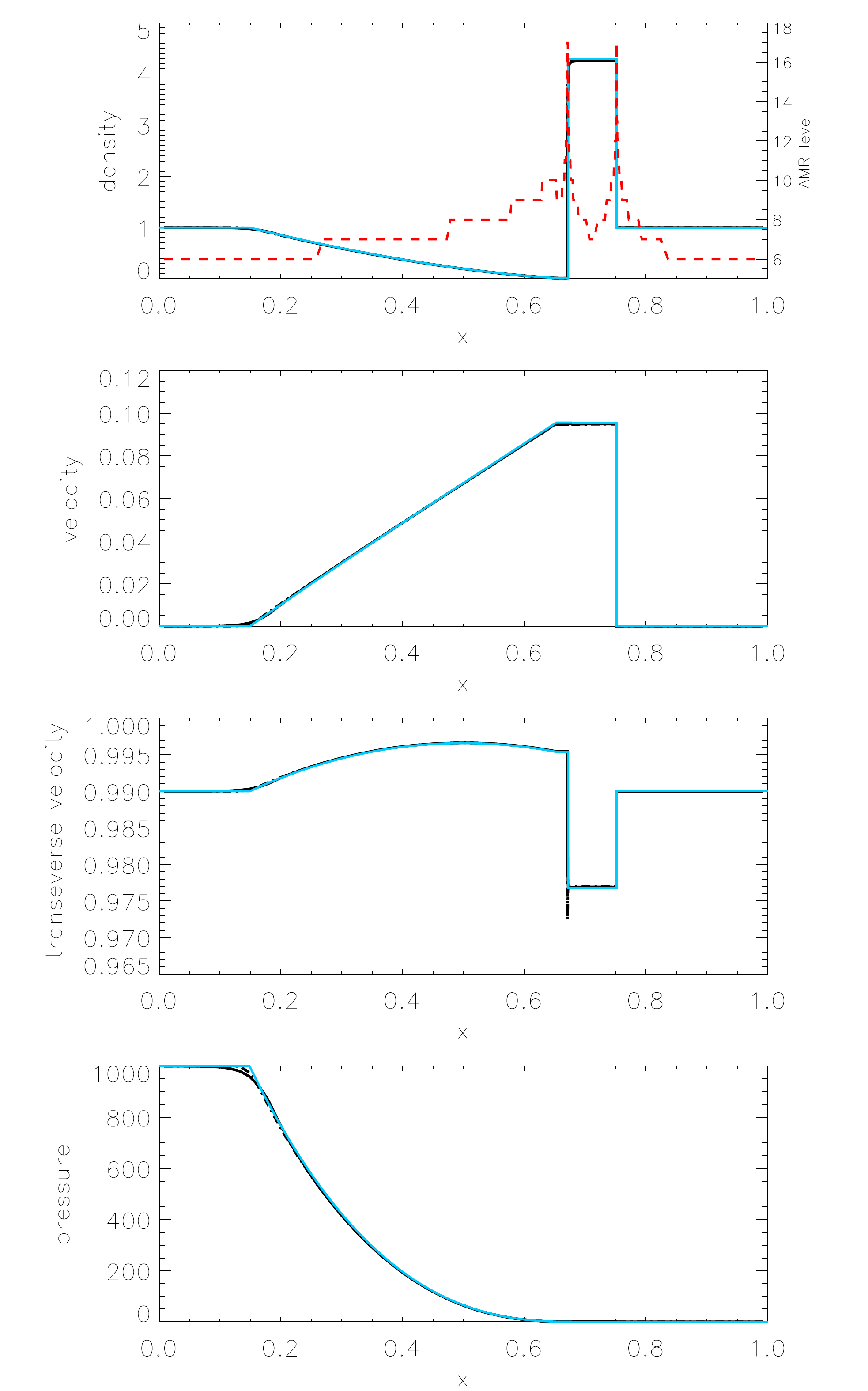}  
  \caption{Density, AMR levels (red dashed line),  parallel and transverse velocity and pressure in the frame of the laboratory for the shock test with $\Gamma_{\rm max}=120$. Left panel\,:\,uniform grid $n_x=512$, middle panel\,:\, uniform grid $n_x=2^{17}=131072$, right panel\,:\,$n_x=64$, $l_{\rm max}=17$. The black solid line represents the MUSCL method, while the black dashed line represents the PLM method. At high resolution, they give the same result, which is very close to the analytic solution.}
  \label{fig:shock23_ryu}
\end{figure*}

 Fig.\,\ref{fig:shock23_ryu} gives the density, parallel and transverse velocity and pressure at $t_{\rm end}=1.8$ for tests with different resolutions. The thick black line represents the simulation using the MUSCL scheme with the \textit{minmod} limiter while the dashed line represents the simulation following the PLM method with the \textit{moncen} limiter. Both methods lead to similar results.  The AMR levels are superposed to the density maps. Refinement is based on the gradients of both the Lorentz factor and pressure. The first simulation has a uniform resolution of 512 cells, which is close to the resolution used by \citet{2005ApJS..160..199M}. Our simulation is similar to theirs, with the shock running ahead of its theoretical position and some inaccuracy in the orientation of the velocity vector. These discrepancies reduce with increasing resolution. In the second simulation, the resolution is also uniform but set to $2^{17}=131 072$ cells. This is the same resolution as \citet{2006ApJS..166..410R} and we obtain a very good agreement with their results and the analytic solution. At high resolution, there is no clear difference between the PLM and MUSCL method. However, such high resolution is prohibitive in multidimensional simulations. The last panel shows a simulation with AMR, the coarse level is set by $n_x=64$ while the highest level $l_{max}=17$. The result is in very good agreement with the expected solution.  The density difference between our results and the analytic solution is lower than one percent, even at the discontinuities.  At the end of the test, the gain of computing time is about a factor 200 compared to the uniform grid. A simulation with the same computing time, but on a uniform grid would have a resolution of slightly more than 4096 cells. A test simulation with a uniform resolution $n_x=4096$ showed a 40$\%$ discrepancy in the density at the beginning of the shocked region, with respect to the analytic solution.  AMR proves to be very helpful tool in the modeling of highly relativistic flows. 

\subsection{2D Inclined shock tube}
2D tests are important to verify the accuracy of the coupling between the different components of the four-vectors.   We performed the above Sod tests  inclining the interface between the two media by  $\theta=21.7^{\circ}$ with respect to a vertical line. The value of $\theta$ is chosen arbitrarily, to avoid any peculiar alignement with respect to the numerical grid, as could occur for $\theta=45^{\circ}$ \citep{2012arXiv1206.6502R}. As the initial separation between both sides is inclined, for a given $y$, the initial conditions are shifted by a few cells with repect to the row of cells just below. The boundary conditions along the $y$ axis follow the same shift. At the bottom boundary one has
\begin{equation}
\left\{
 \begin{aligned}
   \mathbf{U}(i,1)&=\mathbf{U}(i-n_{shift},n_y-2)\\
   \mathbf{U}(i,2)&=\mathbf{U}(i-n_{shift},n_y-1)\\
   \mathbf{U}(i,3)&=\mathbf{U}(i-n_{shift},n_y)
 \end{aligned}
\right.
\end{equation}
for all the cells with $i-n_{shift}=i-(n_y \times tan\theta)>0$. The boundary conditions are periodic along the $x$ axis.  The simulations took 2200 CPU hours.

\begin{figure*}
  \centering
  \includegraphics[width = .3\textwidth ]{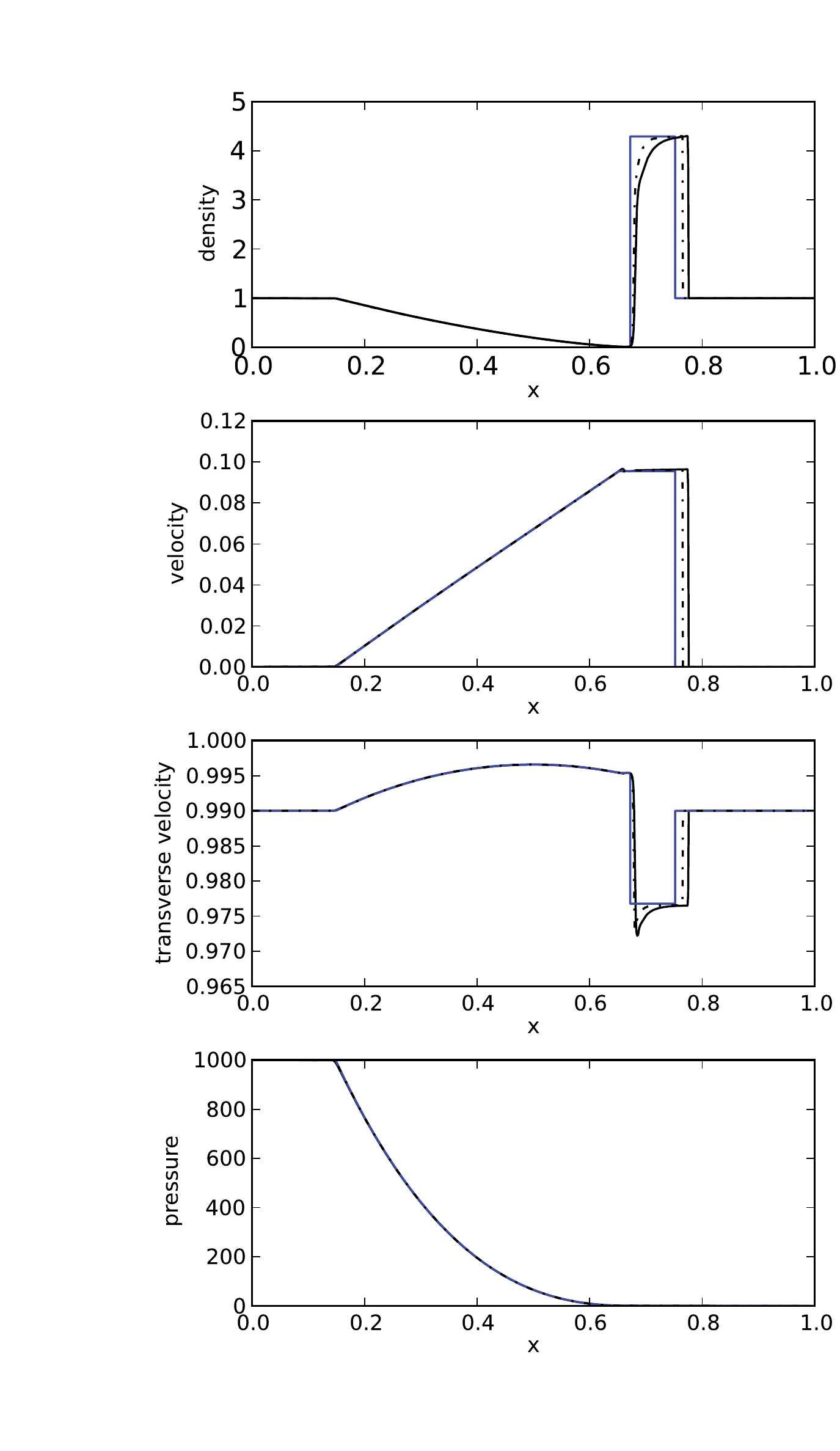}
  \caption{Inclined shock test : Density, velocity, pressure and Lorentz factor along the shock normal. The thick solid lines give the 2D results, the dot-dashed  lines the 1D results from a simulation with the same resolution and the blue lines gives the analytic solution. Note that the resolution is lower than for the 1D test in \S\ref{sec:1D_sod}.}
  \label{fig:2D_sod}
\end{figure*}

The simulations should give a comparable result to 1D simulations with the same resolution.  Fig.\,\ref{fig:2D_sod} shows the different variables in the direction  normal to the shock for the second test using a $12800\times 6400$ uniform resolution. The given values were obtained performing a bilinear interpolation of the 2D si2006ApJS..166..410Rmulation. We overplot the analytic solution and the result from a 1D simulation with the same resolution along the interface between the flows ($n_y/\cos \theta=6400$ cells).
The 2D simulation differs somewhat from its equivalent 1D simulation. This suggests that direct comparison between unidimensional and multidimensional simulations is not possible and that, for high Lorentz factors,  multidimensional simulations need a higher resolution to be numerically accurate. 

In both the 1D and the 2D simulations, the position of the shock is ahead of its theoretical position. 1D and 2D tests have shown this effect weakens at higher resolution.

\subsection{3D jet}\label{sec:jet}

\begin{figure}
  \centering
  \includegraphics[width = .4\textwidth ]{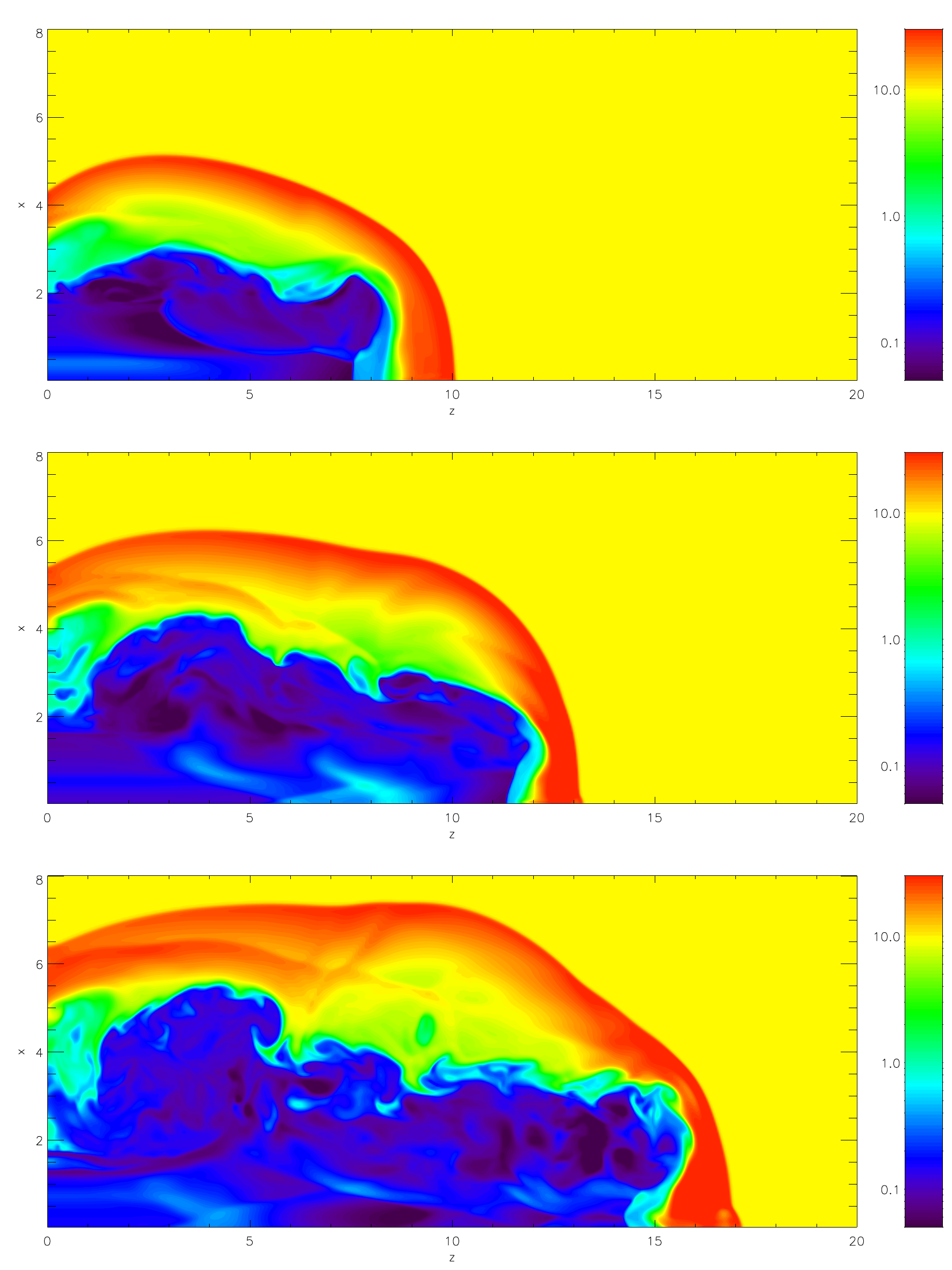}  

  \caption{ Simulation of the propagation of a  3D relativistic jet ($\Gamma_{\rm max}=7.1$). From top to bottom: density at $t=20,30,40$ in a 3D jet starting from the left boundary of the domain.}
   \label{fig:jet_delzanna}
\end{figure}

Relativistic jets have become a widespread means to test multidimensional RHD codes. Their complex dynamics  show the impact of transverse velocities and display the development of instabilities. In this case, no analytic solution exists and validation is done by comparison with former results. We follow the setup by \citet{2002A&A...390.1177D} 
\begin{equation*}
 \left\{
    \begin{array}{llll}
        (\rho,v_x, v_y, v_z ,P) &=& (0.1,0,0,0.99,0.01) &  r\leqslant 1, z\leqslant 1 \\
        (\rho,v_x,v_y,v_z ,P) &=& (10,0,0,0,0.01) &  \mbox{outside} \\
    \end{array}
\right.
\end{equation*}
The length scale is given by the initial radius of the jet $r_0=1$, the size of the box is $20 r_0$, with $l_{\rm min}=6$ and $l_{\rm max}=9$. This gives an equivalent resolution of 25 cells per radius, while the original test was performed with a resolution of 20 cells per radius.  We perform a 3D simulation, while most jets are simulated using axisymmetric 2D simulations \citep{2004A&A...428..703L}.  Except at the injection of the jet, we use outflow boundary conditions. The maximum Lorentz factor is 7.1. The evolution of the  density profile is given on Fig.\,\ref{fig:jet_delzanna}.  One can see the typical features of relativistic jets : the bow shock with the external medium, the cocoon of shocked medium, the relativistic beam and the Mach disk.  The Kelvin-Helmholtz instability develops at the interface between the shocked external medium and the shocked material from the jet.  The global shape is similar to the simulation by \citet{2002A&A...390.1177D}  although the simulation with RAMSES-RHD presents a small extension at head of the jet. This is due to the carbuncle instability \citep{carbuncle} which arises when cylindrical or spherical phenomena are simulated on a Cartesian grid.   Performed on 16 processors, this simulation took 320 mono-CPU hours. The same simulation on a $512^3$ fixed grid took about 500 hours. Due to the complex structure inside the jet, AMR is not very efficient in this type of simulation \citep{2006ApJS..164..255Z}.  

\begin{acknowledgements}
This work was supported by the European Community via contract ERC-StG-200911. Calculations have been performed at CEA on the DAPHPC cluster and using HPC resources from GENCI- [CINES] (Grants 2012046391-2013046391). The authors thank B. Giacomazzo for sharing his code to determine analytic solutions of relativistic shock tubes.
\end{acknowledgements}

\bibliographystyle{aa}
\bibliography{biblio_relat}

\end{document}